\documentclass[conference]{IEEEtran}
\IEEEoverridecommandlockouts

\usepackage{cite}
\usepackage{bbm}
\usepackage{float}
\usepackage{times}
\usepackage{graphicx}
\usepackage{subfigure}
\usepackage{amssymb}
\usepackage{amsmath}
\newtheorem{definition}{Definition}
\newtheorem{problem}{Problem}

\usepackage{slashbox}
\usepackage{booktabs}
\usepackage{array}
\usepackage{multirow}
\usepackage{CJK} 
\usepackage{algorithm}
\usepackage{algorithmic}

\usepackage{threeparttable}
\usepackage{enumerate}

\usepackage{url}
\usepackage[colorlinks,linkcolor=black,anchorcolor=black,citecolor=blue,urlcolor=black]{hyperref}

\ifCLASSINFOpdf
\else
\fi

\begin{document}
%

\title{Expert Finding in Community Question Answering: A Review}


\author{\IEEEauthorblockN{
Sha Yuan\IEEEauthorrefmark{1},
Yu Zhang\IEEEauthorrefmark{2},
Jie Tang\IEEEauthorrefmark{1}\IEEEauthorrefmark{4}\thanks{\IEEEauthorrefmark{4}Correspondence: Tsinghua University, Haidian District, Beijing, China. E-mail: jietang@tsinghua.edu.cn.}
and
Juan Bautista Cabot$\grave{a}$\IEEEauthorrefmark{3}
}
\IEEEauthorblockA{\IEEEauthorrefmark{1}Knowledge Engineering Lab, Department of Computer Science and Technology, Tsinghua University}
\IEEEauthorblockA{\IEEEauthorrefmark{2}Institute of Medical Information, Peking Union Medical College, Chinese Academy of Medical Sciences}
\IEEEauthorblockA{\IEEEauthorrefmark{3}Computer Science Department, University of Valencia}
}

\maketitle

\begin{abstract}
The rapid development recently of Community Question Answering (CQA) satisfies users’ quest for professional and personal knowledge about anything. In CQA, one central issue is to find users with expertise and willingness to answer the given questions.
Expert finding in CQA often exhibits very different challenges compared to traditional methods. Sparse data and new features violate fundamental assumptions of traditional recommendation systems.
This paper focuses on reviewing and categorizing the current progress on expert finding in CQA. We classify all the existing solutions into four different categories: matrix factorization based models (\textbf{MF-based models}), gradient boosting tree based models (\textbf{GBT-based models}), deep learning based models (\textbf{DL-based models}) and ranking based models (\textbf{R-based models}). We find that \textbf{MF-based models} outperform other categories of models in the field of expert finding in CQA. Moreover, we use innovative diagrams to clarify several important concepts of ensemble learning, and find that ensemble models with several specific single models can further boosting the performance. Further, we compare the performance of different models on different types of matching tasks, including \emph{text vs. text}, \emph{graph vs. text}, \emph{audio vs. text} and \emph{video vs. text}. The results can help the model selection of expert finding in practice. Finally, we explore some potential future issues in expert finding research in CQA.
\end{abstract}


\IEEEpeerreviewmaketitle

\section{Introduction}
With the increasing demand of knowledge sharing services, Community Question Answering (CQA) websites, such as Quora, Toutiao and Zhihu, have already obtained the popularization use in reality. It is common to post questions and answers on CQA websites, where users' quest for professional and personal knowledge in various domains can be satisfied. The central task of CQA is to find appropriate users with willingness and relevant expertise to provide high-quality answers for given questions.
This problem has been extensively studied in the past decade. Related research includes expert finding for community-based questions~\cite{Riahi2012Finding, Zhao2016Expert}, expertise modeling~\cite{Han2016Distributed}, and even a comprehensive survey~\cite{Balog2012Expertise}.
Though this problem has been studied before~\cite{Liu2005Finding}, the willingness of experts has been often ignored. This problem becomes more and more seriously -- more than half of the questions on Quora only have one or even do not have any answers\footnote{https://www.quora.com/What-percentage-of-questions-on-Quora-have-no-answers.}.

Expert finding in CQA have generated huge impact to society. It provides a platform to connect questions with experts who can contribute quality answers. Questions about anything can be solved by crowdsourcing in CQA.
For example, CQA can help to find a mathematician for a chef with a math problem. At the same time, cooking tips from the chef will be returned to the mathematician if necessary.
However, it is often hard for CQA to establish such high-quality expert finding.
How to match the questions with interested users' expertise? Can we predict who are the most likely to answer the given questions and what is the probability?
Confronting these challenges, the focuses of expert finding in CQA have changed in practice.

Traditional expert finding problem focused on expert finding~\cite{Riahi2012Finding} and expertise ranking~\cite{Zhao2016Expert}. The experts would be found for the given question based on text matching. In recent years, the core value of the problem is not finding expert, but solving problems by crowdsourcing.
Moreover, expert finding in CQA often exhibits very different challenges compared to traditional methods. The characteristics of expert finding in CQA is summarised as follows.

First, \textbf{crowdsourcing}. The complex and intellectively demanding problems in CQA requires considerable effort and quality contribution. Crowdsourcing is channeling the experts' desire to solve a problem and then freely sharing the answer with everyone. In CQA, the answer of the given question would be obtained by crowdsourcing from a large, relatively open and often rapidly-evolving group of interested experts.

Second, \textbf{sparse data}. The known question and answer pairs are rare compared to traditional expert finding applications. On one hand, seekers spend more time on finding the answer of their question. On the other hand, experts need to answer multiple versions of the same question. This also makes it challenging to directly use a supervised learning approach due to the lack of training samples.

Third, \textbf{new features}. The willingness of expert, the historical behavior of expert, and the quality of answer, all these new features have got more attention. They may contribute to further improve the rationality and effectiveness of expert finding in CQA. For example, the expert who often provides answers with high quality is more likely to answer the similar kinds of questions. How to use these features effectively is widely acknowledged as new challenge that can improve the performance of expert finding in CQA.

Despite of the above challenges, once such expert finding in CQA is successfully formed, its impact is usually tremendous.
Based on these observations, most well-known CQA websites and competitions, such as Quora, Toutiao and Kaggle are striving to match questions with interested users' expertise, that is, to find the best respondents to the questions.
As for this study, we have got the labeled datasets of the competition ByteCup\footnote{https://biendata.com/competition/bytecup2016/.} organized by Toutiao. And therefore we will take these datasets of Toutiao as an example to review the methodologies for expert finding in CQA in the following parts of this paper.

In this paper, we firstly review all the existing expert finding solutions in CQA and classify all the solutions into different categories, including matrix factorization based models (\textbf{MF-based models}), gradient boosting tree based models (\textbf{GBT-based models}), deep learning based models (\textbf{DL-based models}) and ranking based models (\textbf{R-based models}).
In addition, we illustrate the results of all the aforementioned categories of single models on the local validation dataset in the ByteCup competition, and specify the single model obtaining the best performance. The ensemble strategies of the Top $5$ teams who won the competition are also analyzed. We use innovative diagrams to clarify several important concepts of ensemble learning. This work will significantly help the correct understanding and proper use of ensemble learning in practice. Further, we investigate the performance of different models on different types of matching tasks. Finally, we statistically analyze the results of all expert finding solutions in CQA, and summarize the work of this paper.

The remainder of the paper is organized as follows. 
In the next section, we first give a general overview.
Sections~\ref{sec:MF},~\ref{sec:GBT},~\ref{sec:DL} and~\ref{sec:R} present the MF-based models, GBT-based models, DL-based models and R-based models, respectively. Section~\ref{sec:ensemble} specifies the details of ensemble learning.
Section~\ref{sec:result}, \ref{multi} and \ref{discussion} present the results and the corresponding analysis.
Finally, Section~\ref{conclusion} concludes the paper.



\section{Overview}\label{Preliminaries}

\subsection{A Brief History of Expert Finding}
Inspired by recent advances in information management systems, expert finding has attracted a lot of attention in the information retrieval (IR) community~\cite{Lin2017A}.
The core task of expert finding is to identify persons with relevant expertise for the given topic.
Massive efforts have been taken to improve the accuracy of experts finding~\cite{Zhang2007Expert}\cite{Rode2017Entity}\cite{Rafiei2015A}\cite{Boeva2017Data}\cite{Wang2013ExpertRank}\cite{Dargahi2017Skill}\cite{Li2015Social}\cite{Zhou2012Topic}\cite{Liu2013Integrating}. Most existing methods for expert finding can be classified into two groups, including the \emph{authority-based methods}~\cite{Yeniterzi2014Constructing}\cite{Zhu2014Ranking}\cite{Bouguessa2008Identifying}\cite{Liu2011Competition}\cite{Zhou2014An}, which are based on the link analysis of the past expert-topic activities, and the \emph{topic-based methods}~\cite{Zhang2008A}\cite{Deng2009Formal}\cite{Daud2010Temporal}\cite{Topicmodeling2013Topic}\cite{Liu2013An}\cite{Lin2013Finding}\cite{Hashemi2013Expertise}\cite{Yang2013CQArank}, which are based on the latent topic modeling techniques. Moreover, the emerging deep learning models are integrated with aforementioned methods to further improve the performance of expert finding~\cite{Wei2017Collaborative}\cite{Li2017Deep}\cite{Ying2016Collaborative}.
They are capable of effectively learning high dimensional representations of expert information, topic information and expert-topic interactions.

Expert finding has been researched in various areas such as academic~\cite{Rani2015Expert}, organizations~\cite{Karimzadehgan2009Enhancing}\cite{DawitYimam2003Expert}, social networks~\cite{Bozzon2013Choosing}\cite{Kardan2011Expert}\cite{Li2013A}, and more recently question answering communities~\cite{Cheng2015Exploiting}.
Finding experts with relevant expertise for a given topic has potential in many applications in these areas such as finding appropriate reviewers for a paper~\cite{Mimno2007Expertise}\cite{Liang2016Formal}, finding the right supervisor for a student in academic~\cite{Alarfaj2013Finding} and finding the appropriate experts for the questions in CQA~\cite{Li2015A}.

CQA websites, which provide users with a platform to share their experience and knowledge, are very popular in recent years. Successful CQA websites include general ones (such as Toutiao, Quora and Zhihu), and domain-specific ones (such as Stack Overflow). Finding persons with relevant expertise for a specific question in CQA can increase the quality of answers and further improve the crucial problems facing by CQA, such as the low participation rate of users, long waiting time for answers and low quality of answers~\cite{On2017Dynamicity}. Expert finding in CQA is a challenging task which may due to the sparsity of the CQA data, and the emerging features. A great amount of studies have been conducted on expert finding in CQA~\cite{Zhou2012Topic}\cite{Liu2015ZhihuRank}\cite{Riahi2012Finding}\cite{Zhao2015Cold}\cite{Zhao2016Expert}. Before we present the categorization of expert finding techniques, we first describe the notations and definitions used in this paper.

\begin{table*}[htp]
\centering
\begin{threeparttable}
\caption{Performance of different categories of models on different types of matching tasks.}
\vspace{5px}
	\label{tableper}
\begin{tabular}{|p{4cm}|p{1.7cm}<{\centering}|p{1.7cm}<{\centering}|p{1.7cm}<{\centering}|p{1.7cm}<{\centering}|}
		\hline
  		\backslashbox{Model category\kern-2em}{\kern-2em data type} &text VS text &graph VS text &audio VS text &video VS text\\
  		\hline
        MF-based models &${\surd}$ & & &\\
        \hline
  		DL-based models & &${\surd}$ & &${\surd}$\\
        \hline
        GBT-based models & &${\surd}$ &${\surd}$ &\\
        \hline		
        R-based models &${\surd}$ & & &\\
  		\hline
	\end{tabular}
    \begin{tablenotes} 
        \item \centering ${\surd}$ means that this category of models perform the best on that type of data.
    \end{tablenotes}
\end{threeparttable}
\end{table*}

\subsection{Problem Definition}
We present required definitions and formulate the problem of expert finding in CQA. Our goal is to find experts to solve a given question in CQA in the way of crowdsourcing. More specifically, given certain question, one needs to find who are the most likely to $1$) have the expertise to answer the question and $2$) have the willingness to accept the invitation of answering the question.

\vspace{2mm}
\begin{definition}
    \textbf{Expert} is the user with sufficient expertise for a certain \textbf{question} in CQA. The \textbf{expertise} are implied in relevant user documents, social interactions, past activities or personal information of each expert.
\end{definition}
\vspace{2mm}

Given a set of $M$ questions $Q = \left \{q_1,\cdots ,q_M\right \}$, we need to predict which experts $E = \left \{e_1,\cdots ,e_N\right \}$ are more likely to answer these questions.
For simplicity, we reserve special indexing letters for distinguishing experts from questions, where $u$, $v$ represent experts, and $i$, $j$ represent questions.

\vspace{2mm}
\begin{problem}\label{problem1}
    For a given question $i$ and its candidate expert $u\in E$, one needs to predict the \textbf{probability} $\hat{r}_{ui}$ of the expert $u$ answering the question $i$.
\end{problem}
\vspace{2mm}

The ($u,i$) pairs for which $r_{ui}$ is known are stored in the set $\L =\left \{ \left ( u,i \right )|r_{ui}~is~known \right \}$.
The probability $r_{ui}\in \left [ 0,1 \right ]$, high values mean stronger preference of the expert $u$ to answer the question $i$. $\hat{r}_{ui}$ is the predicted probability that the question $i$ will be answered by the expert $u$ based on the labeled data.
Here, it is a supervised learning problem to make prediction with the given labeled data. We need to infer a function from the labeled training examples, and then use the function to label the unknown data. In order to get the function, we need to reduce the error between $\hat{r}_{ui}$ and $r_{ui}$.
Consequently, the objective optimization function is
\begin{equation}
L=\sum l(\hat{r}_{ui},r_{ui})
\end{equation}
where $l$ is the loss function.

Overfitting always happen. If we have too many features, the learned hypothesis may fit the training set very well, but fail to generalize the new examples. There are often two options to solve overfitting. The first is to reduce the number of features. The details is dependent on the specific problem. The second is regularization, which is used to reduce magnitude or values of each feature with parameter $\theta$. It often works well when there are a lot of features, and each of them contributes a bit to the prediction $\hat{r}_{ui}$.

For example, if we use L$2$-norm for regularization,
the optimization problem is transformed into the following problem:
\begin{equation}
\mathbf{\Theta^{*}}=\underset{\arg\min}{\Theta}\sum(l(\hat{r}_{ui},r_{ui})+\sum_{\theta \in \Theta}\lambda _{\theta}\theta ^{2} )
\end{equation}
where $\lambda _{\theta}$ is the regularization coefficient of parameter $\theta$ used in the hypothesis function. As it grows, regularization becomes heavier. Then, we need to find an appropriate optimization method to solve this optimization problem. In this way, we get the parameters of the prediction model, which can be used to label the unknown data.

Typical data in CQA implies large interaction between experts and questions. For instance, some experts prefer to answer than others, and some questions are more likely to be answered than others. In order to account for these affects, it is customary to adjust the data with baseline.

\vspace{2mm}
\begin{definition}
    The \textbf{baseline} for the prediction $\hat{r}_{ui}$ is denoted by $b_{ui}$:
    \begin{equation}\label{eqbase1}
    b_{ui}=\mu+b_u +b_i,
    \end{equation}
\end{definition}
in which, the overall average probability is denoted by $\mu$; the parameters $b_u$ and $b_i$ indicate the observed average deviations of expert $u$ and question $i$, respectively. For example, suppose that we want to get a baseline for the probability of the question $i$ answered by the expert $u$. The average probability over all questions $\mu=0.6$. The expert $u$ tends to answer question lower than the average with probability $0.3$, so $b_u=0.3-0.6=-0.3$. The question $i$ tends to be answered with probability $0.7$, so $b_i=0.7-0.6=0.1$. Thus, the baseline for question $i$ answered by expert $i$ is $b_{ui}=0.6-0.3+0.1=0.4$.


\subsection{Categorization of Expert Finding Techniques}
Based on the survey of possible solutions, we categorize the techniques of expert finding in CQA under four subsettings, including MF-based models, GBT-based models, DL-based models and R-based models. Table.~\ref{tableper} shows the cases where the different approaches are used.

As shown in the Table.~\ref{tableper}, we summarize the performance of these models on different types of matching tasks to explore the scope of application\footnote{More details of experiment results will be clarified in Section~\ref{multi}.}. In the table, \emph{text VS text} means to match text labels with text data, \emph{graph VS text} means to match text labels with graph data, \emph{audio VS text} is to match text labels with graph data; \emph{video VS text} is to match text labels with video data.

We come to the conclusion that MF-based models usually achieve the best performance in the situation of \emph{text VS text}, while DL-based models are rarely used in these situations and not performing well due to the severe sparsity of the text datasets. In addition, R-based models have significant performance in the situation of \emph{audio VS text}, DL-based models often achieve the best in the situation of both \emph{graph VS text} and \emph{video VS text}, which may due to their outstanding power of capturing high dimensional features from graphs and videos. We will discuss these four category solutions in detail below. In addition, ensemble learning of these models will also be discussed.

\section{Matrix Factorization Based Models}\label{sec:MF}
Matrix factorization (MF)~\cite{koren2009mf}, which is a common technique for collaborative filtering (CF)~\cite{Linden2003Amazon}, covers a wide range of applications in recommender system with its variants.
The \emph{Problem~\ref{problem1}} can be modeled as recommendation problem solved by CF, because similar users may answer the similar questions. Therefore, MF can be applied to exploit latent information from data. In this part, we summarize the MF-based models, including MF, Singular Value Decomposition (SVD), SVD++, Bidirection SVD++, Bidirection Asymmetric-SVD (ASVD++) and Factorization Machine (FM).

\subsection{MF}
\begin{figure}[htb]
  \centering
  \includegraphics[width=0.47\textwidth]{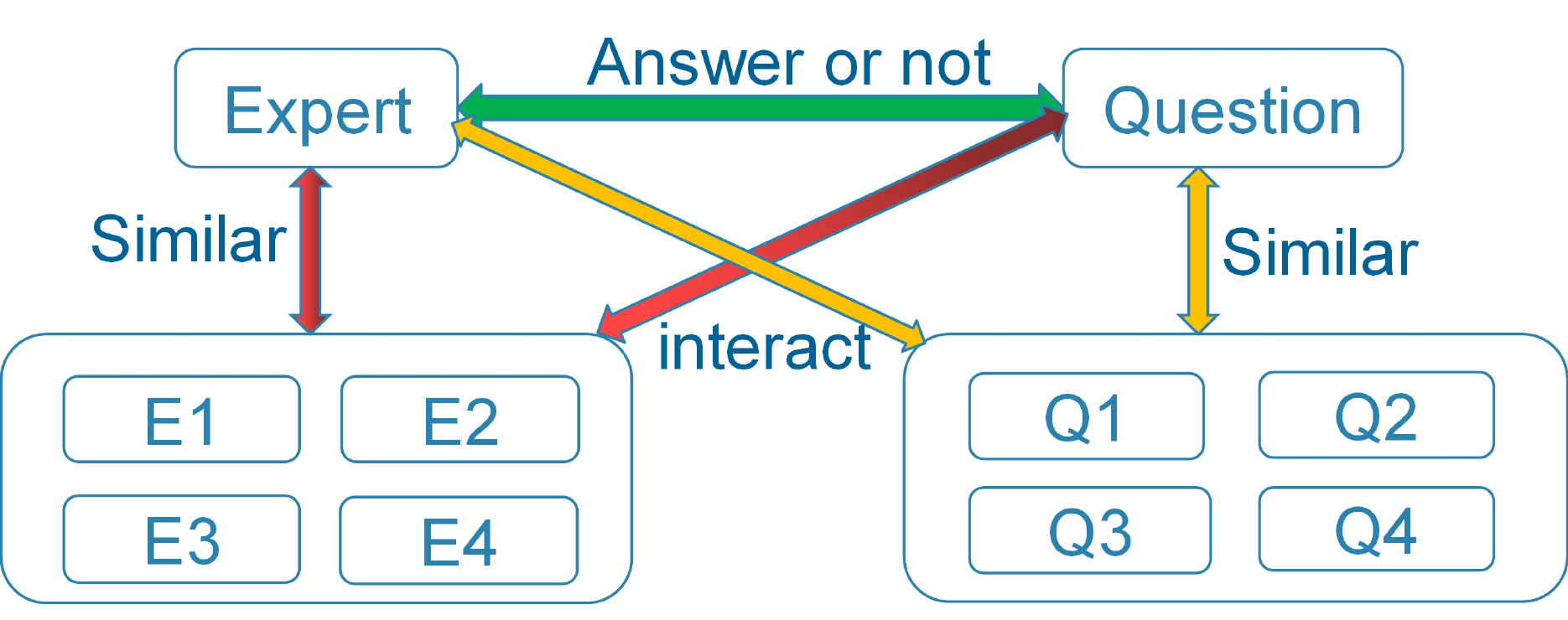}\\
  \caption{Implied Information}
  \label{cf}
\end{figure}

From the application point of view, MF can be used effectively to discover the latent features underlying the interactions between different kinds of entities. For example, several experts have answered same questions before as illustrated in Fig.~\ref{cf}. If some of them (assume the number is $N$) answer a new question, others may also answer the question (assume the probability is $p$). $N$ is larger, $p$ is larger.

From the mathematical point of view, MF is used to factorize a matrix obviously as its name suggesting. The original matrix can be represented by the multiply of two (or more) simple matrices with lower dimension. Let $U$ and $D$ be the set of experts and questions, respectively. Let $\mathbf{R}$ be the record matrix of the expert-question pairs.
If we would like to discover $k$ latent features, we need to find two matrices $\mathbf{P}$ (a $|U|\times k$ matrix) and $\mathbf{Q}$ (a $|D|\times k$ matrix) such that their product approximates $\mathbf{R}$:
\begin{equation}
\hat{\mathbf{R}}=\mathbf{P}^{T}\times \mathbf{Q}\approx \mathbf{R}.
\end{equation}
Thus, matrix factorization maps experts and questions to a joint latent factor space of dimensionality $k$. Each row of $\mathbf{P}$ would represent the strength of the associations between a expert and the features. Similarly, each row of $\mathbf{Q}$ would represent the strength of the associations between a question and the features.

Matrix factorization maps experts and questions to a joint latent factor space of dimensionality $k$, such that expert-question interactions are modeled as inner products in that space.
The resulting dot product $p_u^{T}q_i$ captures the interaction between expert $u$ and question $i$.
\begin{equation}\label{eqmf1}
\hat{r}_{ui} = p_u^{T}q_i.
\end{equation}
Then we directly model the observed probabilities only, while avoiding over-fitting through a regularized model. To learn the factor vectors $p_u$ and $q_i$, the system minimizes the regularized squared error on the set of known probabilities:
\begin{equation}
\min_{P,Q}\sum_{(u,i)\in \L}(r_{ui}-q_i^Tp_u)^2+\lambda(\left \| p_u \right \|^2+\left \| q_i \right \|^2)
\end{equation}
where aforementioned $\L$ is the set of the $(u,i)$ pairs for which $r_{ui}$ is known.

\subsection{SVD}
One benefit of the matrix factorization approach to collaborative filtering is its flexibility in dealing with various data and other application-specific requirements. Eq.~(\ref{eqmf1}) tries to capture the interactions between users and questions without taking the baseline into consideration. Here we combine Eq.~(\ref{eqbase1}) and Eq.~(\ref{eqmf1}) as follows:
\begin{equation}\label{eqsvd}
\hat{r}_{ui}=b_{ui}+p_u^T q_i
\end{equation}
The system learns by minimizing the squared error function, and avoids over-fitting through an adequate regularized model:
\begin{equation}
\min_{P,Q,B}\sum_{(u,i)\in \L}(r_{ui}-\hat{r}_{ui})^2+\lambda(\left \| p_u \right \|^2+\left \| q_i \right \|^2+b_u^2+b_i^2)
\end{equation}

\subsection{SVD++}
MF and SVD models only consider explicit feedback which comes from the interaction between a user and a question. However, we can also obtain implicit feedback from the training data. For instance, a user prefers those questions that he answers in the past. Recommender systems can use implicit feedback to gain insight into user preferences. Indeed, we can gather behavioral information regardless of the user's willingness to provide explicit ratings.
Here, we try to integrate both explicit feedback and implicit feedback. We could get more accurate results by a direct modification of Eq.~(\ref{eqsvd}):
\begin{equation}
\hat{r}_{ui}=b_{ui}+q_i^T(p_u+\left | N(u) \right |^{-\frac{1}{2}}\sum_{j\in N(u)}y_j)
\end{equation}
where $N(u)$ is the set of questions that user $u$ has received invitation. A user $u$ is modeled as $p_u+\left | N(u) \right |^{\frac{1}{2}}\sum_{j\in N(u)}y_j$. $p_u$ is learnt from the given explicit ratings and $\left | N(u) \right |^{\frac{1}{2}}\sum_{j\in N(u)}y_j$ represents the perspective of implicit feedback.
Here, a new set of item factors are necessary, where question $j$ is associated with $y_j \in \mathbb{R}^f$.
Model parameters are learnt by minimizing the squared error function.
\begin{equation}
\min_{P,Q,B,Y}\sum_{(u,i)\in \L}(r_{ui}-\hat{r}_{ui})^2+\lambda\left \| \theta \right \|^2
\end{equation}
where $\theta$ represents the parameters of the model.
SVD++~\cite{Koren2008FM} does not offer the benefits of having less parameters, conveniently handling new users and readily explainable results. This is because the model does abstract each user with a factors vector. However, SVD++ is clearly advantageous in terms of prediction accuracy than SVD.

\subsection{Bidirection SVD++ (SVD\#)} \label{SVD++}
Appending another part of implicit feedback to the original SVD++ model, a new model named bidirection SVD++ model (also called SVD\#) is built. The formula of this model turns to be:
\begin{equation}
    \begin{aligned}
        \hat{r}_{ui} &= b_{ui} + (q_i + \left | R(i) \right |^{-\frac{1}{2}} \sum_{j \in R(i)} x_j)^T\cdot \\
        &(p_u+\left | N(u) \right |^{-\frac{1}{2}}\sum_{j\in N(u)}y_j)
    \end{aligned}
\end{equation}
$R(i)$ is the set of users who answer question $i$.
Here, each question $j$ is associated with $x_j,y_j \in \mathbb{R}^f$.
The other parts of the formula are the same as original SVD++ model.

This model shows the power of representing user/question embeddings using the neighborhood question/user embeddings. However, the embeddings here are static and indepent of time. When the time information is available, a more powerful proposed in ~\cite{Dai2016Recurrent} will be helpful. This method incorporates the embedding co-evolving idea with time series models. The evolution of each user/question embedding depends not only on its old embeddings, but also the embeddings of question/user it interacting with.

\subsection{Bidirection ASVD++}
\label{best_model}
As mentioned in \cite{Koren2008FM},  instead of providing an explicit parameterization for users, users can be represented through the items that they prefer. This model named ``Asymmetric-SVD''(ASVD) offers several benefits: (1) fewer parameters; (2) handle new users; (3) explainability; (4) efficient integration of implicit feedback. Combining the ``bidirection'' strategy described in Sec.~\ref{SVD++}, there is a new model named bidirection ASVD++ model. The formula is listed as below:
\begin{equation}
    \begin{aligned}
        \hat{r}_{ui} &= b_{ui} + (\left | R(i) \right |^{-\frac{1}{2}} \sum_{j \in R(i)} x_j)^T\cdot \\
        &(p_u+\left | N(u) \right |^{-\frac{1}{2}}\sum_{j\in N(u)}y_j)
    \end{aligned}
\end{equation}

\subsection{Factorization Machine}
FM~\cite{Rendle2011FM} is a generic approach based on matrix factorization to mimic most factorization models.
libFM~\cite{Rendle2012libFM} proposed by Steffen Rendle is a software implementation for factorization machines.
It combines the generality of feature engineering with the superiority of factorization models in estimating interactions between variables of large domain.
FM model has the following advantages. Firstly, variable interactions are embedded in the FM model. Secondly, it is able to reliably estimate parameters under very high sparsity. Thirdly, the equation, which depends only on a linear number of parameters, can be computed in linear time. Forthly, it can be applied to a variety of prediction tasks, including regression, binary classification and ranking. In essence, FM model is a matrix factorization based machine learning model and it is similar to linear regression model. We all know the linear regression model has the following formula:
\begin{equation}
\hat{y}(x)=w_0+w_1x_1+...+w_nx_n=w_0+\sum_{i=1}^{n}w_ix_i.
\end{equation}
where $x_i$ is the feature and $\hat{y}$ is the predicted value.

On the basis of model above, if we consider the feature combination, the formula will be changed to the following form:
\begin{equation}
\hat{y}(x)=w_0+\sum_{i=1}^{n}w_ix_i+\sum_{i=1}^{n-1}\sum_{j=i+1}^{n}w_{ij}^{'}x_ix_j.
\end{equation}
Because the sparsity of the feature, we find that many $w_{ij}^{'}$ will be zero after the training. Thus, in order to reduce the number of parameters, FM models the problem by the following formula:
\begin{equation}\label{eqfm}
\hat{y}(x)=w_0+\sum_{i=1}^{n}w_ix_i+\sum_{i=1}^{n-1}\sum_{j=i+1}^{n}(V_i^TV_j)x_ix_j,
\end{equation}
where $V_i$ is the latant vector of the $i^{th}$ feature. We consider a maximum likelihood problem with Eq.~(\ref{eqfm}). To avoid over-fitting, we add some regularization terms. That is, we solve the following optimization problem for FM model.
\begin{equation}
\min_{W,V}\sum_{i=1}^{n}(y_ilog(\sigma(\hat{y_i}))+(1-y_i)log(1-\sigma(\hat{y_i})))+\frac{\lambda}{2}\left \| \theta \right \|^2
\end{equation}
where $\theta$ represents the parameters of the model and $\sigma(x)$ is the sigmoid function.
The learning algorithm of FM mainly contains~\cite{Rendle2012libFM}: Stochastic Gradient Descent (SGD), Alternating Least Squares (ALS) and Markov Chain Monte Carlo (MCMC).



\section{Gradient Boosting Tree Based Models}\label{sec:GBT}
Tree ensemble methods are very widely used in practice. Gradient tree boosting is one of them that shines in many applications. The classic gradient boosting tree and its extension are described in \cite{friedman2001}.
XGBoost \cite{XgboostPPT} is a scalable open source system for tree boosting. The impact of the XGBoost has been widely recognized in a number of machine learning and data mining challenges. One who uses the gradient boosting trees, often chooses XGBoost as the implementation of the Gradient Boosting Regression Trees (GBRT) in the application.

A tree ensemble model uses $K$ additive functions to predict the output.
\begin{equation}
\hat{y}_i=\sum_{k=1}^{K}f_k(x_i),f_k\in \mathcal{F},
\end{equation}
where $\mathcal{F}$ is the space of regression trees (also known as CART). The regularized objective function is listed as follows:
\begin{equation}\label{xgbeq1}
\mathcal{L}=\sum_{i}l(y_i,\hat{y}_i)+\sum_{k}\Omega(f_k),
\end{equation}
where $l$ is a loss function that measures the difference between the prediction $\hat{y}_i$ and the target $y_i$.
The second term $\Omega$ penalizes the complexity of the model:
\begin{equation}\label{xgbeq2}
\Omega(f_k)=\gamma T+\frac{1}{2}\lambda \left \| \omega  \right \|^2.
\end{equation}
$T$ is the number of leaves in the tree. Each regression tree contains a continuous score on each leaf, $\omega_i$ is the score on $i$-th leaf.

Since the tree ensemble model in Eq.(\ref{xgbeq1}) includes functions as parameters but not just numerical vectors, it cannot be optimized using traditional optimization methods such as stochastic gradient descent (SGD) in Euclidean space. In XGBoost, Eq.(\ref{xgbeq1}) is trained in an additive manner.
\begin{equation}
\hat{y}^{(t)}_i=\sum_{k}f_k(x_i)=\hat{y}^{(t-1)}_i+f_t(x_i),
\end{equation}
where $\hat{y}^{(t)}_i$ is the prediction of the $i$-th instance at the $t$-th iteration.
Then, the objective function is:
\begin{equation}
\mathcal{L}=\sum_{i}l(y_i,\hat{y}^{(t-1)}_i+f_t(x_i))+\sum_{k}\Omega(f_k).
\end{equation}
Consider square loss and take Taylor expansion approximation of the loss, we get:
\begin{equation}\label{xgbeq3}
\begin{aligned}
\mathcal{L}^{(t)}&\simeq \sum_{i}[l(y_i,\hat{y}^{(t-1)}_i)+g_if_t(x_i)+\frac{1}{2}h_if^2_t(x_i)]\\
&+\Omega(f_k)+constant,
\end{aligned}
\end{equation}
where
\begin{equation}
g_i=\partial_{\hat{y}^{(t-1)}}l(y_i,\hat{y}^{(t-1)}),
\end{equation}
and
\begin{equation}
h_i=\partial^2_{\hat{y}^{(t-1)}}l(y_i,\hat{y}^{(t-1)}).
\end{equation}

Combining Eq.(\ref{xgbeq1}) and Eq.(\ref{xgbeq3}), we remove constants and get:
\begin{equation}
\mathcal{L}^{(t)}\simeq \sum_{i}[g_if_t(x_i)+\frac{1}{2}h_if^2_t(x_i)]+\gamma T+\frac{1}{2}\lambda\sum_{j}\omega^2_j,
\end{equation}
This is One Variable Quadratic Equation of $\omega_j$. We can compute the optimal weight $\omega ^{*}_j$ of leaf $j$ by
\begin{equation}
\omega ^{*}_j=-\frac{\sum_{i}g_i}{\sum_{i}h_i+\lambda},
\end{equation}
and calculate the corresponding optimal objective function value by
\begin{equation}
\tilde{\mathcal{L}}^{(t)}=-\frac{1}{2}\sum_{j}\frac{(\sum_{i}g_i)^2}{\sum_{i}h_i+\lambda}+\lambda T,
\end{equation}
In practice, the greedy algorithm, that starts from a single leaf and iteratively adds branches to the tree, is usually used for evaluating the split candidates.
It is impossible to efficiently do the exact greedy algorithm when the data does not fit entirely into memory. And then, the approximate algorithm for split finding is proposed in XGBoost. More details can be found in \cite{Xgboost}.

\section{Deep Learning Based Models}\label{sec:DL}
Recently, deep learning models have been widely exploited in various matching tasks with remarkable performance. Applying deep learning models into recommender system has been gaining momentum due to its state-of-the-art performances on popular benchmarks for recommender systems, such as MovieLens\footnote{https://grouplens.org/datasets/movielens/1m/.} and Netflix challenge datasets. Among those deep learning based recommender systems, an autoencoder based system "AutoRec" and a neural autoregressive based system "CF-NADE" have been utilized in the \emph{Problem~\ref{problem1}}. Moreover, a semantic matching model named Match-SRNN, which can model the recursive matching structure between experts and questions, has been also used before.

\subsection{Autoencoder Model} \label{subsec: autorec}
AutoRec~\cite{Autorec2015} is an autoencoder based collaborative filtering model. Similar to traditional CF, AutoRec has two variants: an user-based autoencoder and an item-based autoencoder. They can respectively take user partial vectors and item partial vectors as input, project them into a hidden layer to learn the lower-dimensional representations, and further reconstruct them in the output layer to predict missing ratings for the purpose of recommendation.

While AutoRec is used in the \emph{Problem~\ref{problem1}}, experts are regarded as users, questions as items, and the question distribution data as rating matrix. The question distribution data consists of question push notification records that indicate whether the expert answered the question (if answered, the tag is $1$; otherwise $0$). Then the AutoRec model is deployed to predict the ratings of unknown expert-question pairs.

Both user-based AutoRec and an item-based AutoRec are exploited in expert finding in CQA. Experiment results show that item-based model performs better which may be due to the higher variance of user partial vectors. However, item-based AutoRec is not performing well than MF-based models as before. The reason may be that the dataset of Toutiao is more sparse than the MovieLens dataset.

\subsection{Neural Autoregressive Model}
Inspired by the Restricted Boltzmann Machine (RBM) based CF model, an emerging Neural Autoregressive Distribution Estimator (NADE) based CF model named CF-NADE~\cite{CFNADE} is proposed. It can model the distribution of expert ratings. CF-NADE with only one hidden layer can defeat all the previous state-of-the-art models on recommendation tasks upon the MovieLens $1$M, MovieLens $10$M and Netflix datasets. Furthermore, CF-NADE can be further extended to a deep model with more hidden layers which can further boost the performance.

CF-NADE, which is designed to model the ordering of the ratings, is a feed-forward and neural autoregressive architecture for CF tasks. Ideally, the order of items should follow the time-stamps of ratings. However, empirical study shows that random drawing permutations for each user also generates favourable performances. Since the expert IDs as well as the question IDs are anonymized and the descriptions of expert and questions in the dataset have been encoded into ID sequences, it is feasible to deploy CF-NADE to this competition without time-stamps information. While training the CF-NADE model, the experts and questions are considered as users and items, and the rating matrix is derived from question push notification records like in Section~\ref{subsec: autorec}. Experiment results show that the performance of CF-NADE model in the \emph{Problem~\ref{problem1}} is similar to the AutoRec model, in which item-based CF-NADE performs better than user-based CF-NADE but still not comparable to the matrix factorization based models such as SVD++ and ASVD++. Moreover, the CF-NADE model, though worth trying, is not integrated into any final ensemble models because it significantly reduces the performance when incorporated into ensemble models.

\subsection{Match-SRNN}
Furthermore, the expert finding problem in CQA can also treated as a text matching problem. Thus, text matching methods can be applied to this task which can take advantage of textual features such as characters and words in the the expert and question descriptions. For the \emph{Problem~\ref{problem1}}, a deep text matching model called Match-SRNN~\cite{Match-SRNN} is applied to model the interaction information between texts to further predict new expert-question pairs. The Match-SRNN model contains three parts: a neural tensor network to capture the character/word level interactions, a spatial recurrent neural network (spatial RNN) applied on the character/word interaction tensor to capture the global interactions recursively, and a linear scoring function to calculate the final matching score. The Match-SRNN model views the generation of the global interaction between two texts as a recursive process which can not only obtain the interactions between nearby words, but also take advantage of long distant interactions.

\section{Ranking Based Models}\label{sec:R}
The evaluation criterion in this task is normalized discounted cumulative gain (NDCG), thus ranking based model is a natural fit for this target. There are two kinds of ranking based models appearing in the expert finding problem in CQA, including ranking based FM and ranking based SVM.
\subsection{Ranking based FM}
The basic idea of this model is coming from the FM method. We modify the objective function to optimize the pair-wise ranking loss. Let $N^+$ denotes the number of positive samples and $N^-$ denotes the number of negative samples. Besides, $x_i$ denotes the negative instances and $x_j$ denotes positive instances. Then we solve the following optimization problem for ranking based FM.
\begin{equation}
\begin{aligned}
\min_{W,V}&\frac{1}{N^++N^-}\sum_{i=1}^{N^-}\sum_{j=1}^{N^+}log(1+exp(\hat{y}(x_i)-\hat{y}(x_j)))\\
&+\frac{\lambda}{2}\left \| \theta \right \|^2
\end{aligned}
\end{equation}
where $\hat{y}(x)$ is the prediction in the Eq.~(\ref{eqfm}). We expect that those positive samples have higher prediction score than those negative samples.

\subsection{Ranking based SVM}
ranksvm~\cite{joac2006}, which is a linear pairwise ranking model, has also been used in the problem. Specifically, we first build the feature vectors for each user-question pair appeared in the training/test sets. Then those training pairs with same questions are organized together as a list. The pairwise constraints are then built within each list. 

\begin{figure*}[t]
\centering
\subfigure[Diagram of Bagging.]{
\includegraphics[width=0.53\textwidth, height=0.18\textheight]{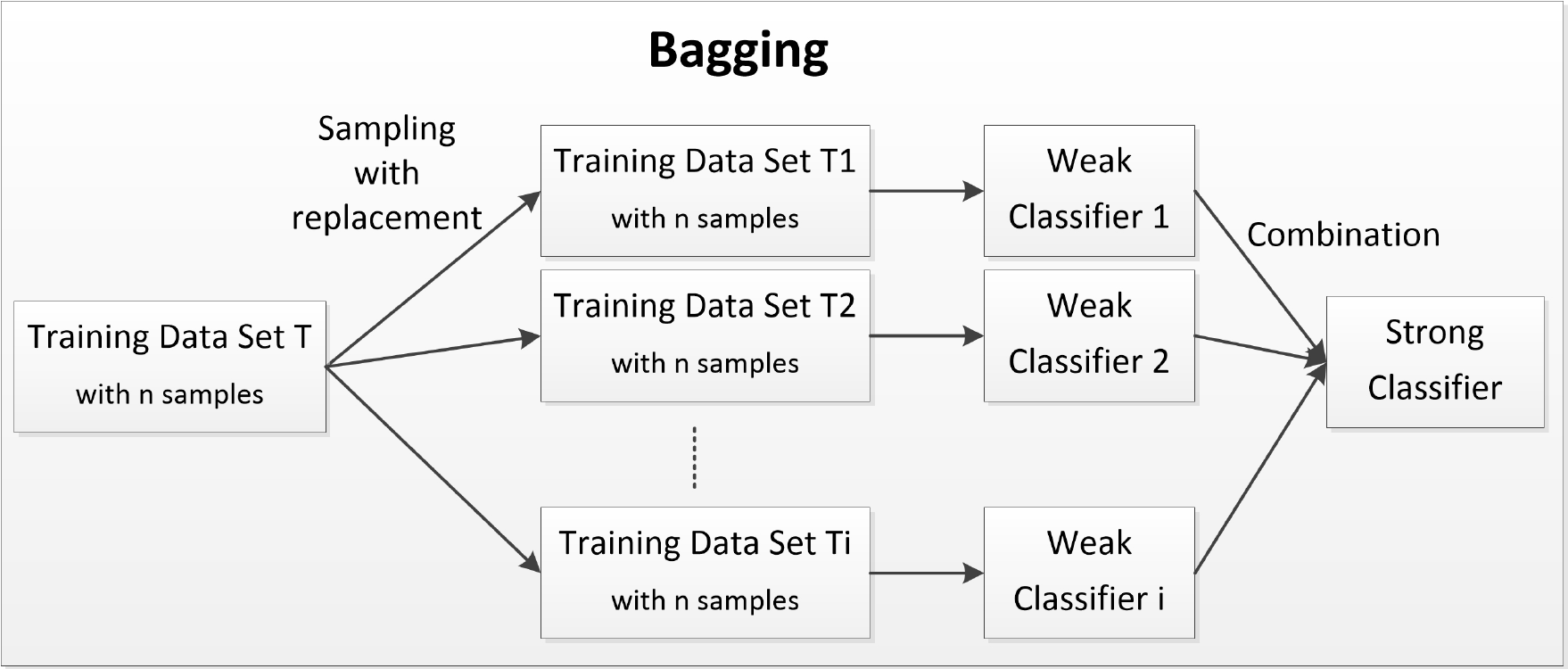}
\label{fig:bagging}}
\subfigure[Diagram of Boosting.]{
\includegraphics[width=0.43\textwidth, height=0.18\textheight]{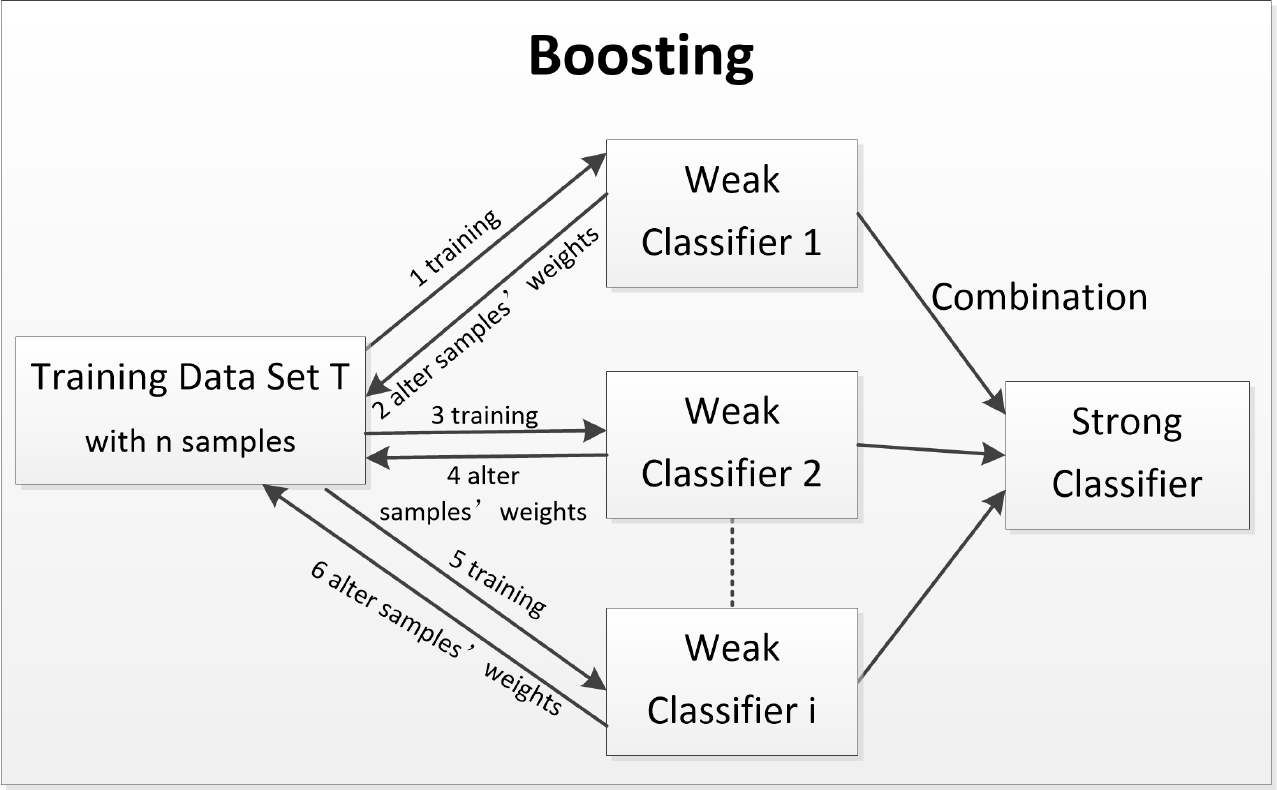}
\label{fig:boosting}}
\subfigure[Diagram of Stacking.]{
\includegraphics[width=0.96\textwidth]{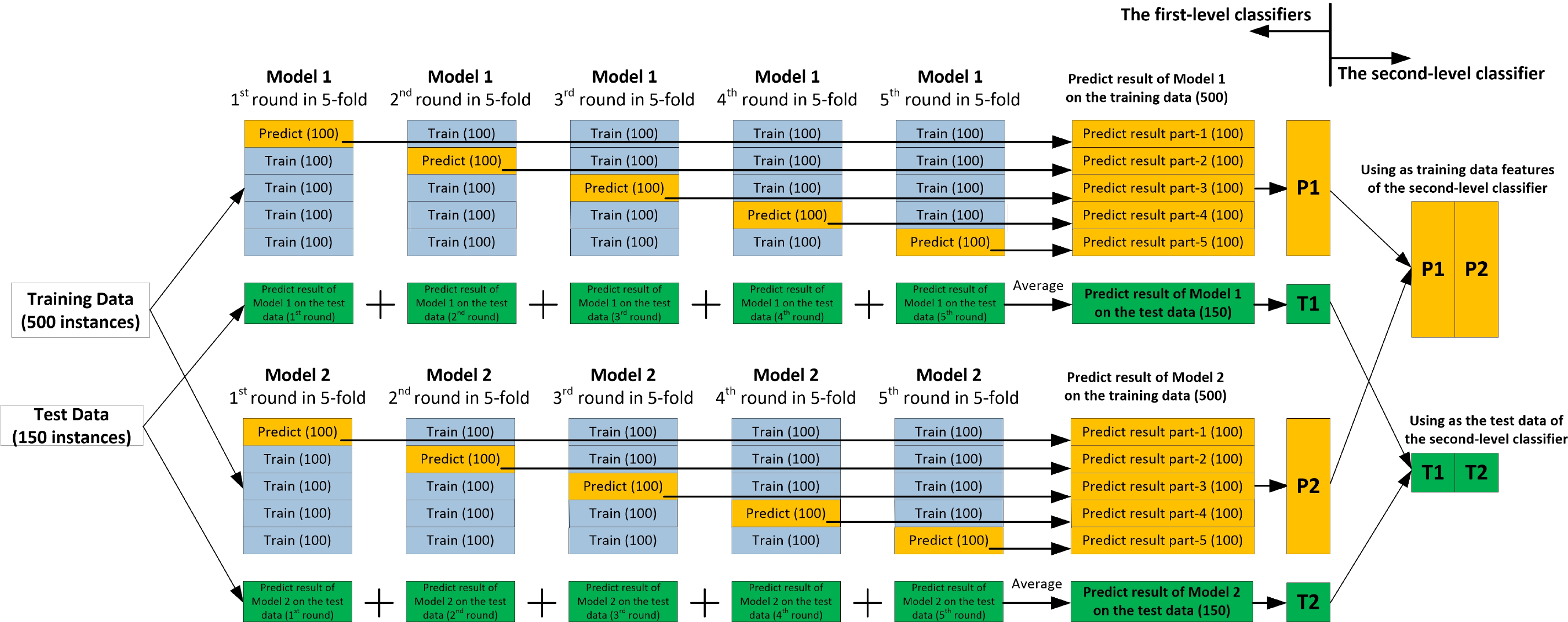}
\label{fig:stacking}}
\caption{Diagram of Ensemble Learning.}
\label{fig:ensemble}
\end{figure*}


\section{Ensemble Learning}\label{sec:ensemble}
During the review of the ensemble learning solutions, we find that many contestants are obscure about the concept of ensemble learning, especially Stacking. These proper nouns are often inappropriately used in ensemble learning. Here, we comb through the relevant concepts of ensemble learning that are widely used in practice.
In machine learning, ensemble learning (also called ensemble method~\cite{zhou2012} before) is a proper noun. It is a method of using multiple learning algorithms to obtain better predictive performance than that could be obtained by any of the component learning algorithms alone.
Ensemble learning can be used for classification problems, regression problems, feature selection, anomaly detection and so on. In the following part, we will use classification as an example.

If we use ensemble learning to improve the overall generalization ability of classifiers, the following two conditions should be satisfied. Firstly, differences exist between the base classifiers. The performance of the ensemble classifier will not be improved, if it is just an ensemble of the same kind of base classifiers. Secondly, the classification accuracy of every base classifier must be larger than $0.5$. If the classification accuracy of the base classifier is less than $0.5$, the classification accuracy of the ensemble classifier will decline with the increasing of ensemble size. If the two aforementioned conditions are satisfied, the classification accuracy of the ensemble classifier will edge up to $1$ with the increasing of ensemble size. Generally, the classification accuracy of a weak classifier is just slightly better than random guess, while a strong classifier can make make very accurate predictions. The base classifiers are referred to as weak classifier.

There are two key points in ensemble learning. How to generate base classifiers with difference? How to combine the results of the base classifiers? We will introduce ensemble learning from these two aspects.

\subsection{Types of Ensemble Learning}
According to how the base classifiers are constructed, there are two paradigms of ensemble learning, the parallel ensemble learning and the sequential ensemble learning. In the parallel ensemble learning, the base classifiers are generated in parallel, with Bagging~\cite{Breiman1996} as a representative. In the sequential ensemble learning, the base classifiers are generated sequentially, with Boosting~\cite{Friedman2000Additive} as a representative.

\subsubsection{\textbf{Bagging}}
Bagging (\textbf{B}ootstrap \textbf{agg}regat\textbf{ing}) was proposed to improve classification accuracy by combining classifiers of randomly generated training sets. Fig.~\ref{fig:bagging} illustrates the diagram of Bagging.

Bagging applies bootstraping~\cite{Johnson2001Bootstrap} to obtain the data subsets for training the base classifiers. In detail, given a training data set containing $n$ training examples, a sample of $n$ training examples will be generated by random sampling with replacement. Some original examples appear more than once, while some original examples are not present in the sample. If we need to train $m$ number of base classifiers, this process will be applied $m$ times. The combination methods used by Bagging are the most popular strategies, that is, voting for classification and averaging for regression.
Here, the final classification results are determined by averaging on the respective results of these classifiers.

\subsubsection{\textbf{Boosting}}
Instead of resampling the training dataset as Bagging does, Boosting adjusts the distribution of the training dataset. Fig.~\ref{fig:boosting} illustrates the diagram of Boosting.
Boosting is an iterative process to generate base classifiers sequentially, where the later classifiers focus more on the mistakes of the earlier classifiers. In each round, the weight of the samples, which have been classified incorrectly, will be increased in the training dataset. The weight of the samples, which have been classified correctly, will be decreased in the training dataset.
Finally, the ensemble classifier is a weighted combination of these weak classifiers.

\subsection{Combination Methods}
The combination method plays a crucial role in ensemble learning.
After generating a set of base classifiers, ensemble learning resorts to combination method to achieve an ensemble classifier with strong generalization ability, rather than trying to find a best single classifier.
Generally, the most popular combination methods used in practice are \textbf{Voting}, \textbf{Averaging} and \textbf{Learning}.
Voting and Averaging are the most popular and fundamental combination methods for nominal outputs and numeric outputs, respectively. These two methods are easy to understand and use. Here, we mainly focus on the Learning, with Stacking (stacked generalization) as a representative.

\subsubsection{Stacking}
Unlike Voting and Averaging, Stacking is a general combining procedure where the base classifiers are combined non-linearly in a serial model. In Stacking, the base classifiers are called the first-level classifiers, while the combiner is called the second-level classifier (or meta-classifier). The basic idea of Stacking is to train several first-level classifiers using the original training dataset. And then, a new dataset generated from the first-level classifier is used to train the second-level classifier, where the outputs of the first-level classifiers are regarded as the input features of the new training dataset, and the original labels are still the labels of the new training data.

In the training phase of Stacking, if all the instances in the training dataset are used to train the first-level classifiers, and the outputs of the first-level classifiers are used to train the second-level classifier, there will be a high risk of over-fitting. Therefore, the instances used for generating the input of the meta-classifier need to be excluded from the training instances of the first-level classifiers. Generally, a cross validation is used to avoid this problem.

\begin{table*}[!htpb]
\centering
\caption{Designed features.}
\vspace{5px}
	\label{table1}
	\begin{tabular}{p{4cm}|p{1.5cm}<{\centering}|p{6cm}|p{1cm}<{\centering}|p{1cm}<{\centering}}
		\hline
  		Name &Notation &Description &Type &+/-\\
  		\hline
  		Anonymized expert user ID &\emph{uID} &The unique identifier of each expert user. & id  & +\\
  		Expert user tags &\emph{uTag} &The tag of user information. & category  & + \\
  		Word ID sequence of user &\emph{uwordIDseq} &Segmented user description. Each word is replaced by a unique wordID. & category  & - \\
  		Character ID sequence of user &\emph{ucharIDseq} &Segmented user description. Each character is replaced by a unique charID. & category  & - \\
  		Anonymized question ID &\emph{qID} &The unique identifier of each question. & id  & + \\
  		Question tag &\emph{qTag} &The tag of each question. & category  & + \\
  		Word ID sequence of question &\emph{qwordIDseq} &Same as uwordIDseq instead of question description. & category & - \\
  		Character ID sequence of question &\emph{qcharIDseq} &Same as ucharIDseq instead of question
description. & category & - \\
  		Number of upvotes &\emph{upvoteNum} &Number of upvotes of all answers to this question. & numeric & + \\
  		Number of answers &\emph{ansNum} &Number of all answers to this question.  & numeric & + \\
        Number of top quality answers &\emph{topAnsNum} &Number of top quality answers to this
question. & numeric &+\\
  		Implicit expert &\emph{imE} &Expert list with implicit relationship. & category & ++ \\
  		Implicit question &\emph{imQ} &Question list with implicit relationship.  & category & ++ \\
  		\hline
	\end{tabular}
\end{table*}

Taking a Stacking model with $2$ first-level classifiers and $5$-fold cross validation as an example, Fig.~\ref{fig:stacking} illustrates the diagram of Stacking. There are $500$ instances in the training dataset. Using the Model $1$ (the first-level classifier) in Fig.~\ref{fig:stacking} as an example, in the $5$-fold cross validation, the training dataset is divided into $5$ parts, and each part has $100$ instances. Four of them (with $400$ instances in total) are used to train the Model $1$. The remaining one part (with $100$ instances) is used to do prediction. The prediction results ($5$ parts with $500$ instances in total) are used as the features of the input of the second-level classifier. In every round in the $5$-fold cross validation, the trained Model $1$ makes prediction on the test dataset (with $150$ instances). After $5$ rounds, there are $5$ parts of the prediction results on the test dataset. Making an average of these $5$ parts, there are still $150$ instances in the final prediction result of Model $1$ on the test dataset.

Generally, Stacking can be viewed as a specific combination method of the Learning combination strategy. What's more, it can also be regarded as a general framework of many ensemble methods used in practice.

\section{Results}\label{sec:result}
In terms of the evaluation criteria, NDCG will be used. Specifically, we will rank the experts based on the forecasted probability for a certain question, and evaluate the $NDCG@5$ and $NDCG@10$ of ranking results. The final evaluation formula is: $NDCG@5*0.5 + NDCG@10*0.5$.

\subsection{Data Analysis}
In this paper, we analyze the problem of expert finding in CQA by taking the data of ByteCup competition as an example. The data provided for the competitors consisting of expert finding records in CQA with three types of information: expert tags, question data and question distribution data:
\begin{enumerate}
\item The expert tag data, which contains IDs of all expert users, their interest tags, and processed profile descriptions.
\item The question data, which contains IDs of all questions, processed question descriptions, question categories, total number of answers, total number of top quality answers, total number of upvotes.
\item The question distribution data: $290000$ records of question push notification, each contains the encrypted ID of the question, the encrypted ID of the expert user and if the expert user answered the question ($0$=ignored, $1$=answered).
\end{enumerate}

The training set, validation set and test set are divided based on these records. The training set is used for the training of the model. Validation set is used for online real-time evaluation of the algorithm. Test set is used for the final evaluation.

All expert ID and question ID are encrypted to protect user privacy.
Also for privacy protection purpose, the original descriptions of the questions and the experts are not provided. Instead, the ID sequence of the characters (each Chinese character will be assigned an ID) and the ID sequence of the words after segmentation (each word will be assigned an ID) are provided. Validation and testing labels have not been published. They are used for online evaluation and final evaluation only.

\begin{table*}[htp]
\centering
\begin{threeparttable}
\caption{Designed features.}
\vspace{5px}
	\label{table2}
\begin{tabular}{|p{1.5cm}<{\centering}|p{0.5cm}<{\centering}|p{0.5cm}<{\centering}|p{0.5cm}<{\centering}|p{0.5cm}<{\centering}|p{1.2cm}<{\centering}|p{1cm}<{\centering}|p{1.2cm}<{\centering}|p{0.5cm}<{\centering}|p{0.5cm}<{\centering}|}
		\hline
  		\backslashbox{Team\kern-2em}{\kern-2em Features} &\emph{uID} &\emph{uTag} &\emph{qID} &\emph{qTag} &\emph{upvoteNum} &\emph{ansNum} &\emph{topAnsNum} &\emph{imE} &\emph{imQ}\\
  		\hline
  		Team-$1$  &$\bullet$ &$\bullet$ &$\bullet$ &$\bullet$ &$\circ$ &$\circ$ &$\circ$ &$\bullet$ &$\bullet$\\
        \hline
  		Team-$2$ &$\bullet$ &$\circ$ &$\bullet$ &$\circ$ &$\circ$ &$\circ$ &$\circ$ &$\bullet$ &$\bullet$\\
        \hline
  		Team-$3$ &$\bullet$ &$\circ$ &$\bullet$ &$\circ$ &$\circ$ &$\circ$ &$\circ$ &$\bullet$ &$\bullet$\\
        \hline
        Team-$4$ &$\bullet$ &$\circ$ &$\bullet$ &$\circ$ &$\bullet$ &$\bullet$ &$\bullet$ &$\bullet$ &$\bullet$\\
        \hline
        Team-$5$ &$\bullet$ &$\circ$ &$\bullet$ &$\circ$ &$\circ$ &$\circ$ &$\circ$ &$\bullet$ &$\bullet$\\
  		\hline
	\end{tabular}
    \begin{tablenotes} 
        \item \centering $\bullet$ means that the feature is used. $\circ$ means that the feature is not used.
    \end{tablenotes}
\end{threeparttable}
\end{table*}

\subsection{Feature Extraction}\label{subSecFeature}
We summarise all possible features in Table~\ref{table1}. 
The expert user tags \emph{uTag} may be multiple tags, i.e., $18$, $19$ and $20$ may represent baby, pregnancy and parenting, respectively. In the feature of \emph{uwordIDseq}, user descriptions (excluding modal particles and punctuation) are first segmented, and then each word will be replaced by the Character ID, i.e., $284$/$42$ may represent ``Don't Panic". In the feature of \emph{ucharIDseq}, user descriptions (excluding modal particles and punctuation) are first segmented, and then each character will be replaced by the Character ID, i.e., $284$/$42$ may represent ``BE".
The question tag \emph{qTag} may be a list of single tags, i.e., $2$ may represent fitness.
The feature \emph{upvoteNum}, \emph{ansNum} and \emph{topAnsNum} may indicate the popularity of the question.

We also study the positive/negative contributions of each feature.
As Table~\ref{table1} illustrated, four features, including \emph{uwordIDseq}, \emph{ucharIDseq}, \emph{qwordIDseq} and \emph{qcharIDseq}, have negative impact on the model performance. The implicit features \emph{imE} and \emph{imQ}, which have strong positive influence on the model performance，are needed to be considered in the prediction model.

Table.~\ref{table2} illustrates the features used by the top $5$ teams in the competition ByteCup. The four features including \emph{uwordIDseq}, \emph{ucharIDseq}, \emph{qwordIDseq} and \emph{qcharIDseq}, that have negative impact on the model performance shown in Sec.~\ref{subSecFeature}, have not been used by any team. Therefore, we does't include them in Table.~\ref{table2}.
Although there are nine positive features, simply combining all of them will not lead to the best performance. All top $5$ teams use the four features, including \emph{uID}, \emph{qID}, \emph{imE} and \emph{imQ}. The latent features \emph{imE} and \emph{imQ} underlying the interactions between different kinds of entities have important influence on the performance.

\begin{figure}[!htbp]
    \centering
    \includegraphics[width=1\linewidth]{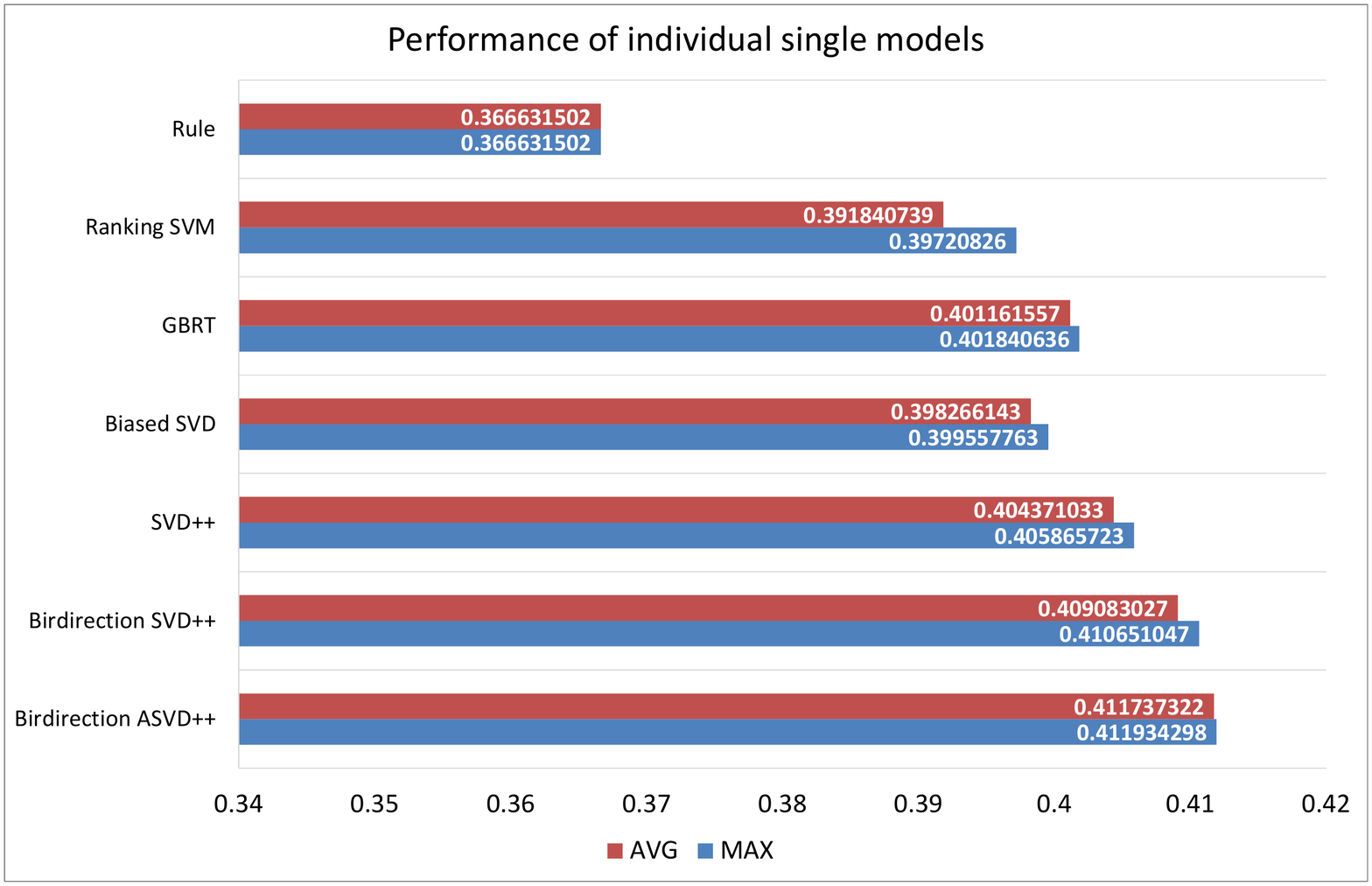}
    \caption{Individual model performances on local validation dataset.}
    \label{fig:single_model}
\end{figure}

\subsection{Results of Single Models}
SVDFeature~\cite{SVDFeature} and Factorization Machine(libFM)~\cite{Rendle2012libFM} tools are used for MF-based models. XGBoost~\cite{Xgboost} is used for GBT-based models. The code based on Theano framework is used for the DL-based models.

The results of all aforementioned categories of single models on the local validation dataset is illustrated in Figure.~\ref{fig:single_model}. From the figure we can see that, some single models such as ASVD and bidirectional SVD++ make good performances. However, there are also weak models such as ranksvm and simple heuristic based method. In general, the MF-based models perform better than others including GBT-based models and DL-based models, which performs well in many other kinds of applications. We used different settings of parameters (max depth of each tree, number of trees, and boosting step size) to train several XGBoost models. Based on the experiments on local validation dataset, the performance of these models (refer to the performance of models starting with ``GBRT'' in Fig.~\ref{fig:single_model}) are reasonable, but not as good as MF-based models. Nevertheless, they do improve the performance of the final ensemble model. These models have quite different objective and underlying assumptions than MF-based methods. Therefore, a decent weak model will still improve the final ensemble results.

\begin{figure}[!htbp]
    \centering
    \includegraphics[width=0.95\linewidth]{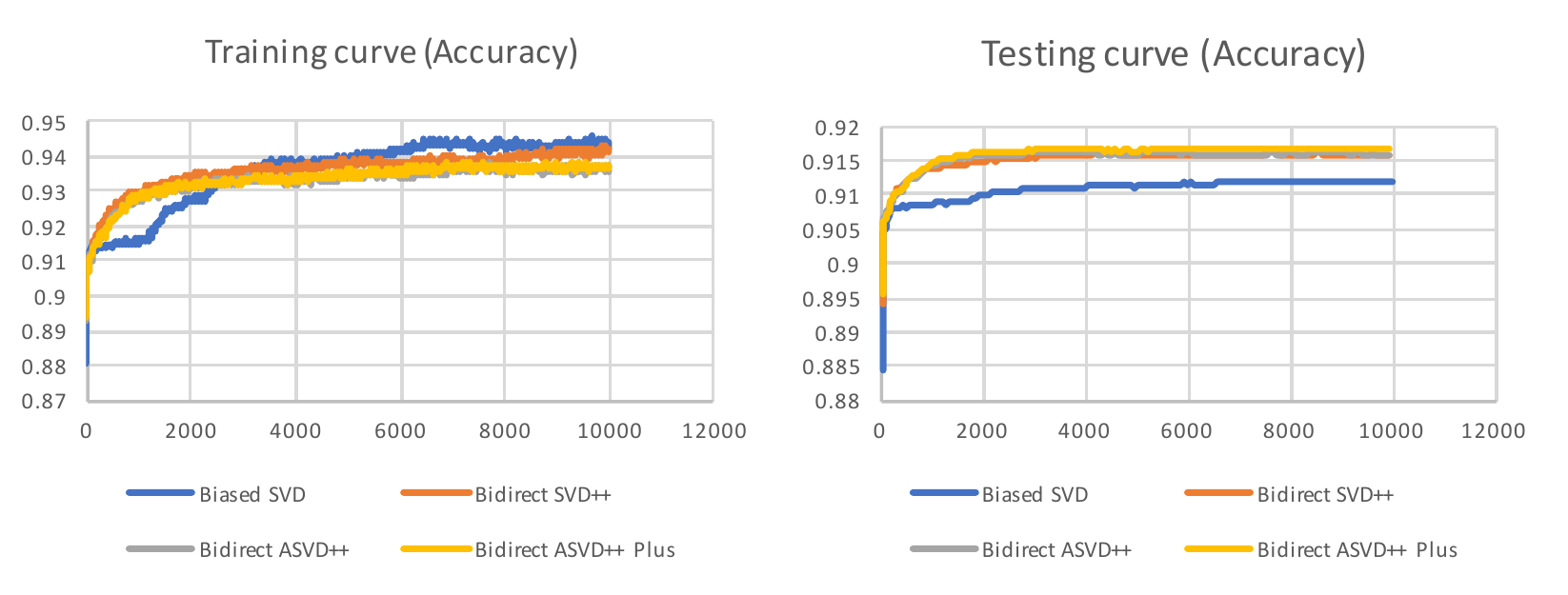}
    \caption{MF-based models training/testing curve. }
    \label{fig:mf}
\end{figure}

In the MF-based models, the bidirection ASVD++ performs the best. What's more, if more implicit information is used, such as ratting action in online validation dataset or online test dataset, the model performance could be further improved. This phenomenon is reflected in Fig.~\ref{fig:mf}. The accuracy of the bidirect ASVD++ is highest, followed by the bidirect ASVD++, the bidirect SVD++ and the bidirect SVD in the descending order.

Table.~\ref{Table.3} illustrates the parameters for the bidirection ASVD++ that achieves the best performance. Markov Chain Monte Carlo (MCMC) is used for the learning method in the model. Table.~\ref{Table.4} illustrates the best performance of the bidirection ASVD++ on the local validation dataset, the online validation dataset and the online test dataset. The results are $0.41193$, $0.52412$ and $0.50551$, respectively.

\begin{table}[htpb]
  \centering
  \caption{Parameters for the bidirection ASVD++.}
  \label{Table.3}
  \begin{tabular}{lc}
    \hline
  	Parameters & Value \\
  	\hline
  	Learning method & MCMC\\
  	\#Factor & 8 \\
  	\#Iteration & 10000 \\
  	Task & binary classification \\
  	Stdev for init. of 2-way factors & 0.1 \\
  	\hline
  \end{tabular}
\end{table}
	
\begin{table}[htpb]
  \centering
  \caption{Performance of bidirection ASVD++.}
  \label{Table.4}
  \begin{threeparttable}
	\begin{tabular}{lc}
	  \hline
  	  Test Set & Performance (nDCG) \\
  	  \hline
  	  Local Validation & 0.41193\\
  	  Online Validation & 0.52412 \\
  	  Online Test & 0.50551${}^*$ \\
  	  \hline
	\end{tabular}
    \begin{tablenotes}
		\item[*] Already rank first among all single models.
	\end{tablenotes}
  \end{threeparttable}
\end{table}

\subsection{Results of Ensemble Models}
Taking the ensemble models of the Top $5$ teams who won the competition ByteCup as the example, we analysis the results of the ensemble models.

\subsubsection{Team-$1$}
As shown in Table.~\ref{table:Ensemble}, Team-$1$ combines $45$ models linearly with different settings (features, tools or hyper-parameters) using the linear ridge regression.  Specifically, they do $5$-fold cross validation on the local validation set. The final ensemble model is trained using local validation set. Note that, the predictions of local validation set are from those models trained on local training set. Thus the training set are not involved in the ensemble step.
They also ensemble the predictions from same model with different parameters, such as different latent dimensions or different objective functions of matrix factorization models. The small variations make the single model more robust.
To avoid the bias due to different scales, they do whitening for each model's prediction before ensemble.

Team-$1$ takes the predictions of each candidate model, and does a linear combination of those predicted values to make the final prediction. The score of these candidate models range from $0.367$ to $0.412$, they tune the weights of them based on the rating prediction on local validation set. The prediction ensemble of a set of base models further improves the performance. Finally, they get the score of $0.50812$ on the final leaderboard.
Team-$1$ has also tried to use nonlinear ensemble method, such as the gradient boosting tree, to do the ensemble. However, they found such tree models are very easy to over-fit the training set. It is also hard to regularize the model to get a good test performance.

\begin{figure}[htbp]
    \centering
    \includegraphics[width=0.95\linewidth]{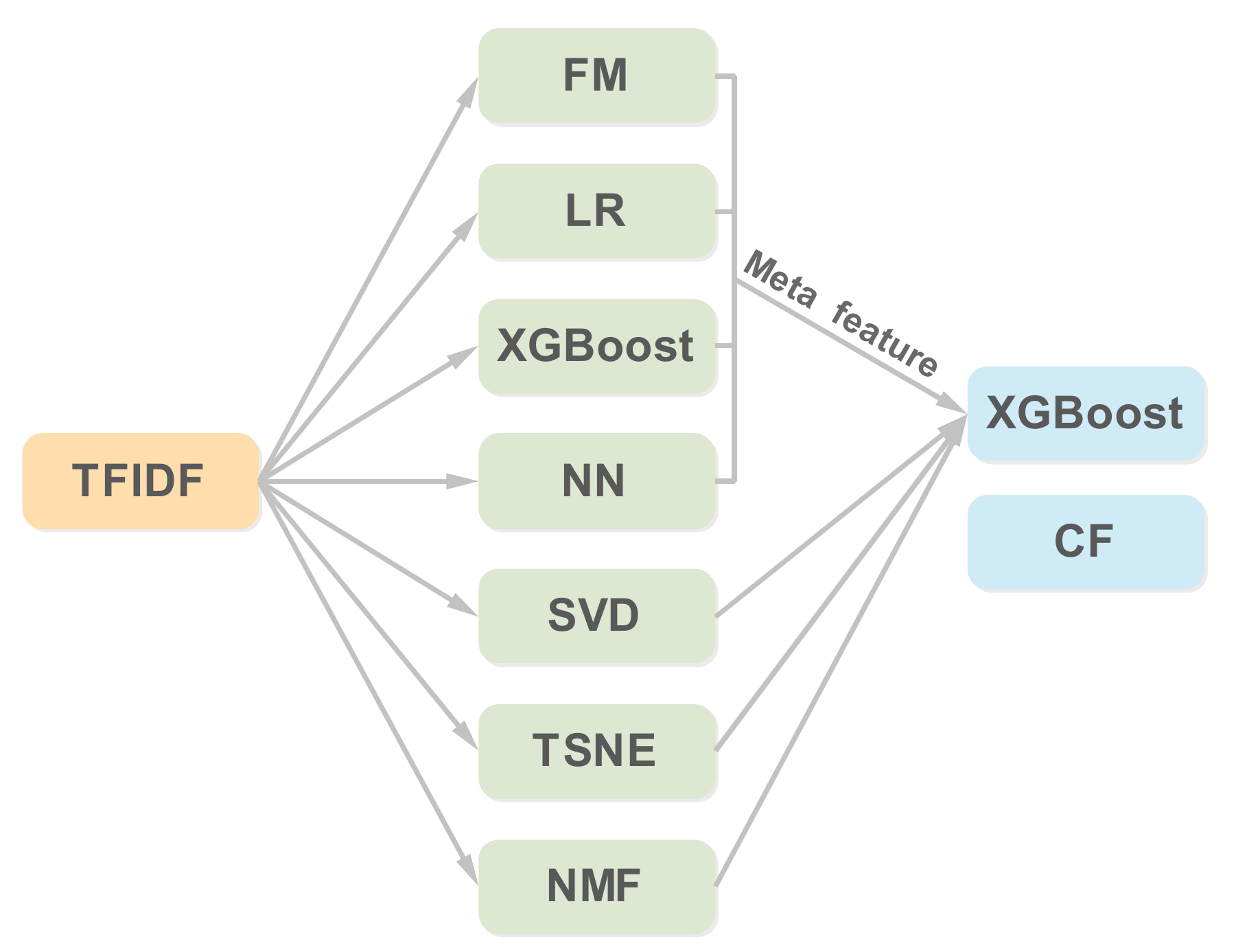}
    \caption{Diagram of Stacking Used by Team $2$}
    \label{fig:tianqiong}
\end{figure}

\begin{table*}[!htpb]
\centering
\begin{threeparttable}
\caption{Ensemble models used by the top $5$ teams.}
\vspace{5px}
	\label{table:Ensemble}
\begin{tabular}{|p{1.5cm}<{\centering}|p{5cm}<{\centering}|p{1.5cm}<{\centering}|p{2.8cm}<{\centering}|}
		\hline
  		\backslashbox{Team\kern-1em}{\kern-1em Method} &Details of the ensemble model &Final results &Compare with Team-$1$ \\
  		\hline
  		Team-$1$  &Linearly combine all models in Fig.~\ref{fig:single_model}  &$0.50812$ &$0$\\
        \hline
  		Team-$2$ &Use Stacking strategy illustrated as Fig.~\ref{fig:tianqiong} &$0.50307$ &$-1\%$\\
        \hline
  		Team-$3$ &FM+CF$~^*$ &$0.49905$ &$-1.82\%$\\
        \hline
        Team-$4$ &MF+CF &$0.49231$ &$-3.21\%$\\
        \hline
        Team-$5$ &FM+RFM+(FM+RFM)+MF+SVD+(SVD++) &$0.49003$ &$-3.69\%$\\
  		\hline
	\end{tabular}
    \begin{tablenotes} 
        \item[*]\centering FM+CF represents the linear weighted sum of FM and CF.
    \end{tablenotes}
\end{threeparttable}
\end{table*}

\subsubsection{Team-$2$}
For every expert, there is a list of questions that have been answered. Here, Team-$2$ regards the expert-question list as a document, and each question is a term. The TF-IDF of each question is calculated and used as the feature \emph{imQ}. Similarly, The TF-IDF of each expert is calculated and used as the feature \emph{imE}.

Team-$2$ uses the method of Stacking to integrate several single models. The Stacking strategy used by them is illustrated in Fig.~\ref{fig:tianqiong}.
In the Stacking, FM, Logistic Regression (LR), XGBoost and Neural Network (NN) are the first-level classifiers. The results of them are used as inputs of the next layer, called meta features. SVD, TSNE~\cite{tsne2017}, NMF~\cite{nmf1994} is used to get the dimension reduction features of the original features. Finally, the meta features and the dimension reduction features are combined to train the XGBoost.

The used NN has one hidden layer, in which the activation function is ReLu(Rectified Linear Units), the droupout rate is $0.75$. Adam~\cite{Adam2014} is also used here to optimize the model. XGBoost is trained in the following steps. They uses the social graph to model the relationship between experts and questions $<E, Q>$. The experts and questions are regarded as nodes in an undirected graph. If a expert is invited to answer a question, there will be an undirected edge between them. DeepWalk~\cite{deepwalk2014} is used to convert $<E, Q>$ to work vector, which then be used to train XGBoost.

In addition, they find three implied messages of CF based on the observation and analysis of the issues and data.
\begin{itemize} 
\item If a expert has accepted most of the invitation for answering question, he will be more likely to accept the new invitation to answer question.
\item Experts have answered some same questions. If some of them (assume the number is $N$) answer a new question, others may also answer the question (assume the probability is $p$). $N$ is larger, $p$ is larger.
\item If questions $Q1$ and $Q2$ are given to the same user, $Q1$ and $Q2$ may be involved in the same field. If $Q1$ is answered by an expert, $Q2$ may be answered by the expert too.
\end{itemize}

And then, they combine the results of Stacking and CF by weight $2:1$. Finally, they get the score of $0.50307$ on the final leaderboard. Only $1\%$ less than Team-$1$.

\subsubsection{Team-$3$}
The weight of the question that is related to the expert \emph{uid} is regarded as the feature \emph{imQ} by Team-$3$. It is calculated as the reciprocal of the question numbers answered by the expert \emph{uid}. The weight of the expert that is related to the question \emph{qid} is regarded as the feature \emph{imE}. It is calculated as the reciprocal of the expert numbers who answer the question \emph{qid}. FM is achieve by libFM.

In CF, the probability of expert answering question is calculated as the weighted sum of the average similarity between experts and the average similarity between questions. The similarity between questions is calculated as the weighted difference between the positive similarity of the question and the negative similarity of the question. The positive similarity of question is the number of experts who have similar behavior on the specific question and answer the test question. The negative similarity of question is the number of experts who have similar behavior on the specific question and not answer the test question. The similarity between experts is calculated similarly as the similarity between questions.

As shown in Table.~\ref{table:Ensemble}, Team-$3$ combines the results of FM and CF with the linear weighted sum. Finally, they get the score of $0.49905$ on the final leaderboard. $1.82\%$ less than Team-$1$.

\subsubsection{Team-$4$}
As shown in Table.~\ref{table:Ensemble}, Team-$4$ combines the results of MF and CF with the linear weighted sum.
In the scheme of CF, the prediction is calculated as the formula shown below:
\begin{equation}
pred(u,i)=\bar{r}_{u}+\frac{\sum_{v\in N(u)}sim(u,i)*(r_{v,i}-\bar{r}_v)}{\sum_{v\in N(u)}sim(u,i)},
\end{equation}
where $sim(u,i)$ is calculated by
\begin{equation}
sim(u,i)=\frac{\sum_{i}(r_{u,i}-\bar{r}_u)(r_{v,i}-\bar{r}_v)}{\sqrt{\sum_{i}(r_{u,i}-\bar{r}_u)^2}\sqrt{\sum_{i}(r_{v,i}-\bar{r}_v)^2}}.
\end{equation}
$N(u)$ is the set of neighbors of the specific expert $u$. The number $n$ of $N(u)$ is hyper parameter needed to be tune. They use $n=5000$ in the final model.

Finally, they get the score of $0.49231$ on the final leaderboard. $3.21\%$ less than Team-$1$.

\subsubsection{Team-$5$}
Team-$5$ combines the results of $6$ individual models on the validation set, including FM, ranking based FM (RFM), the linear weighted sum of FM and RFM, three MF-based models (MF, SVD and SVD++).
Assuming the predictions of the user-question pairs from the $6$ individual models are $pred_1, pred_2, pred_3, pred_4, pred_5, pred_6$, respectively. A weight is assigned to every individual model and the final prediction of the user-question pairs is computed by the following formula:
\begin{equation}
\begin{aligned}
pred&=\alpha_1pred_1+\alpha_2pred_2+\alpha_3pred_3\\
&+\alpha_4pred_4+\alpha_5pred_5+\alpha_6pred_6
\end{aligned}
\end{equation}
After the ensemble, the performance of the model turns out to be better.

What's more, Team-$5$ finds a rule in the training set, and it can be used in the validation set to improve the model performance. In the training set, a certain user-question pair only appears once or twice and a user answers the question once at most. Therefore, they assume that expert won't answer the same question twice and it is consistent with the reality. When the user-question pair appears in the validation set and it also appears in the training set where the user answers the question, they predict that user won't answer the question again. This rule helps to boost the performance on the validation set again.

Finally, they get the score of $0.49003$ on the final leaderboard. $3.69\%$ less than Team-$1$.

\begin{figure*}[t]
\centering
\includegraphics[width=1\textwidth]{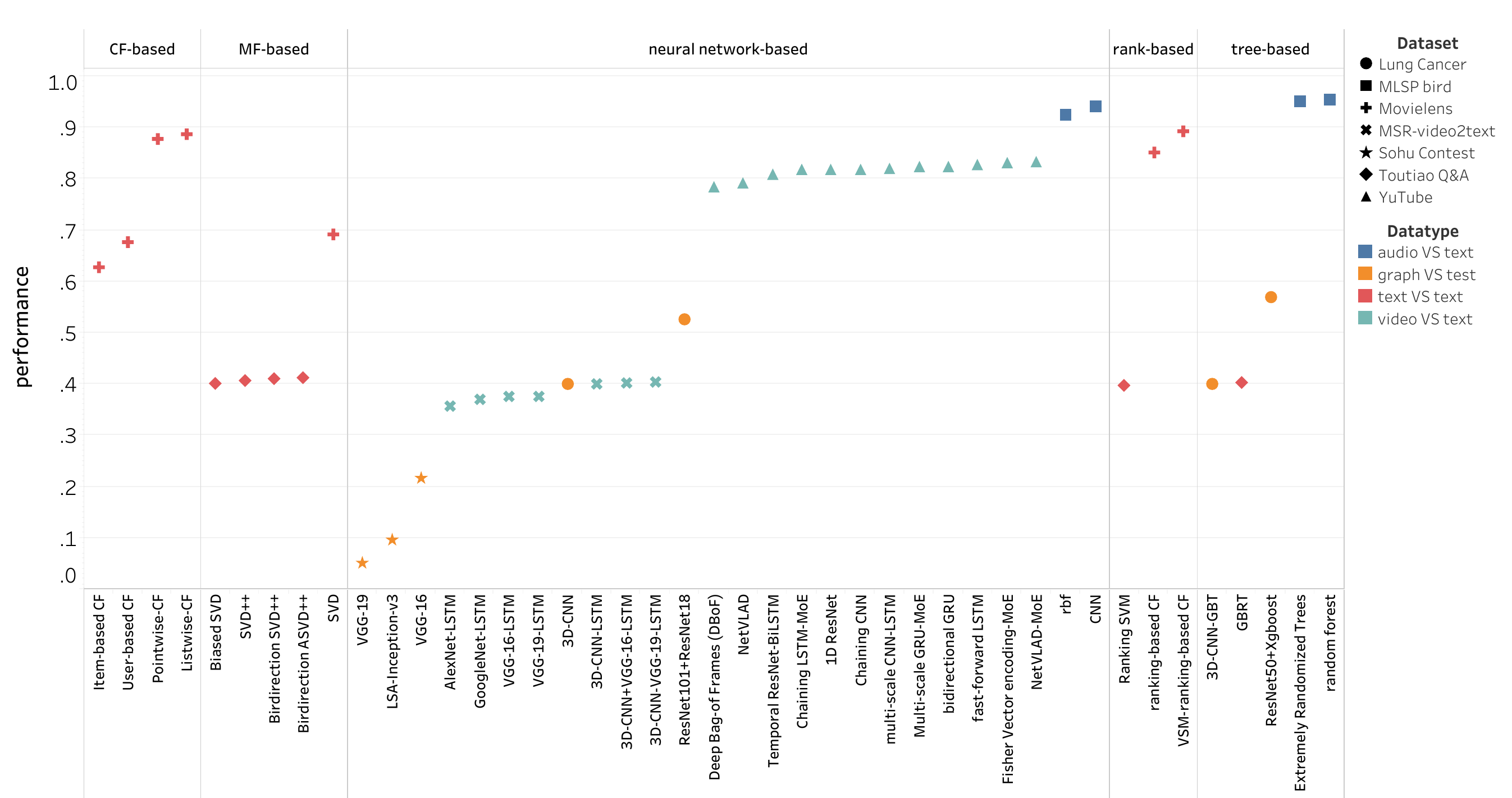}
\caption{Performances of diverse models on different type of datasets.}
\label{fig:multi_methods_dataset}
\end{figure*}

\section{Diverse models on different types of matching tasks}\label{multi}
In this section, we compare the performance of diverse models on different types of matching tasks to explore the difference among the models on different matching tasks (Figure.~\ref{fig:multi_methods_dataset}). Totally, seven matching tasks were involved in the study including:
\begin{enumerate}
  \item Toutiao: The evaluation metric of ByteCup is $NDCG@5*0.5+NDCG@10*0.5$;
  \item Movielens: Movie recommendation on MovieLens data with evaluation metric $NDCG@10$;
  \item Sohu Contest: Sohu Programming Contest\footnote{https://biendata.com/competition/luckydata/.} on news pictures data with evaluation metric average NDCG;
  \item Lung Cancer: Data Science Bowl 2017\footnote{https://www.kaggle.com/c/data-science-bowl-2017.} on Lung CT images data with evaluation metric LogLoss;
  \item MLSP bird: MLSP $2013$ Bird Classification Challenge\footnote{https://www.kaggle.com/c/mlsp-2013-birds.} on bird sounds audio data with evaluation metric micro-AUC;
  \item YouTube: Google Cloud \& YouTube-8M Video Understanding Challenge\footnote{https://www.kaggle.com/c/youtube8m.} on YouTube videos data with evaluation metric Global Average Precision$@20$;
  \item MSR-video$2$text: Video to Language Challenge\footnote{http://ms-multimedia-challenge.com/2016/challenge.} on MSR-video$2$text data with evaluation metric $BLEU@4$.
\end{enumerate}

Based on the data type of the tasks, we classified the seven tasks into $4$ categories. There are : $1$) \emph{text vs. text}, which means to match text labels with text data, includes ByteCup and Movie recommendation; $2$) \emph{graph vs. text}, which means to match text labels with graph data, contains Sohu Programming Contest and Data Science Bowl $2017$; $3$) \emph{audio vs. text}, which aims to match text labels with audio data, includes MLSP $2013$ Bird Classification Challenge; $4$) \emph{video vs. text}, which is to match text labels with video data, includes Google Cloud \& YouTube-$8$M Video Understanding Challenge and Video to Language Challenge.

The models used in the seven tasks are also classified into four categories including MF-based models, GBT-based models, R-based models and DL-based models.
As shown in Figure.~\ref{fig:multi_methods_dataset}, MF-based models and rank-based models are used only in \emph{text vs. text} category of matching tasks, while DL-based models are not employed in these tasks since they are not performing well (which may due to the severe sparsity of the datasets). MF-based models usually achieve the best performance in \emph{text vs. text} category of matching tasks. In addition, DL-based models achieve the best performance in \emph{graph vs. text} and \emph{video vs. text} categories, which may due to their outstanding power of capturing high dimensional features from graph and video, and they are also utilized in the \emph{audio vs. text} category. Finally, GBT-based models have significant performance on \emph{audio vs. text} category.

\section{Discussion}\label{discussion}
In this article, we statistically analyze all the existing solutions for the expert finding problem in CQA. We summarise the results analysis and the learned lessons learned in this part.

\subsection{Results Analysis}
We describe the different individual methods used in the task, and also introduce several types of ensemble learning. And then, we present the results of both of them. It is worth noting that the different individual methods get scores from $0.3665$ to $0.4119$ when used independently. The results of ensemble learning range from a score of $0.49003$ to a score of $0.50812$. Since the data used in the task is the real data from Toutiao with about $580$ million users, even minor improvements can affect millions of users.

Based on the analysis of the solutions and the observation of the results, we find that the ensemble methods outperform any of the single models when they were used independently. That is, ensemble learning really outperforms every single component model, if the two conditions mentioned in Sec.~\ref{sec:ensemble} are both satisfied.
Although there are some model with poor performance, the use of them with other different kind of models leads to a considerable improvement of the prediction. YES! A weak model in combination with other different kind of models can still improve the performance of the final ensemble model. In general, the combination of different kinds of models even with a weak model\footnote{Its accuracy is larger than $0.5$.} leads to significant performance improvements over every single component model.

\subsection{Important Lessons}
As known from the No Free Lunch (NFL) Theorem, none of the algorithms is better than a random one. In the field of machine learning, there isn't an almighty algorithm that is applicable to all situations. Different data sets and different problems have different best algorithms respectively. In previous years, XGBoost shows its absolute advantage in the structured data. However, it puts up a poor show than MF-based models in this task. It is a reasonable explanation that the dataset here is more sparse than movie rating datasets used in previous tasks.

As noticed, a single model won't win. This shows that, as expected, the field of machine learning is getting stronger. This paper witnesses the advantage of ensemble learning applied to the combination of different learning models. In addition, many mobile social platforms in China, such as WeChat, Sina Weibo, Toutiao and so on, have hundreds of million users. Even minor improvements of the solution results can affect millions of users.

Moreover, from the survey of the performance of different models on different types of matching types, we learned that MF-based models and rank-based models are more suitable for \emph{text vs. text} matching tasks, DL-based models and GBT-based models achieve the best results for \emph{audio vs. text} matching tasks. DL-based models are appropriate for both \emph{video vs. text} and \emph{audio vs. text} matching tasks.

\section{Conclusion}\label{conclusion}
This survey paper focuses on the expert finding problem in CQA. Given certain question, one needs to find who are the most likely to $1$) have the expertise to answer the question and $2$) have the willingness to accept the invitation of answering the question.
We have reviewed most existing solutions and classify them to four different categories: MF-based models, GBT-based models, DL-based models and R-based models. Experimental results demonstrate the effectiveness and efficiency of the MF-based models in the expert finding problem in CQA.

In the future, several important research issues need to be addressed. First, how to efficiently integrate the implicit feedback is an open problem. Obviously, implicit feedback becomes increasingly important in practical application, because users provide much more implicit feedback than explicit one. In addition, explainability is usually ignored in the research. The existing methods face real difficulties to explain predictions. Finally, how to make sure that the established model is no needed to be re-trained is a crucial issue in expert finding in CQA.
We hope that the overview presented in this paper will advance the discussion in the expert finding technologies in CQA.

\section*{Acknowledgment}
This work is supported by the National Natural Science Foundation of China ($61561130160$), and the National High Technology Research and Development Program of China ($863$ Program) ($2015AA124102$).



\bibliographystyle{IEEEtran}
\bibliography{reference}

\begin{thebibliography}{10}
\providecommand{\url}[1]{#1}
\csname url@samestyle\endcsname
\providecommand{\newblock}{\relax}
\providecommand{\bibinfo}[2]{#2}
\providecommand{\BIBentrySTDinterwordspacing}{\spaceskip=0pt\relax}
\providecommand{\BIBentryALTinterwordstretchfactor}{4}
\providecommand{\BIBentryALTinterwordspacing}{\spaceskip=\fontdimen2\font plus
\BIBentryALTinterwordstretchfactor\fontdimen3\font minus
  \fontdimen4\font\relax}
\providecommand{\BIBforeignlanguage}[2]{{%
\expandafter\ifx\csname l@#1\endcsname\relax
\typeout{** WARNING: IEEEtran.bst: No hyphenation pattern has been}%
\typeout{** loaded for the language `#1'. Using the pattern for}%
\typeout{** the default language instead.}%
\else
\language=\csname l@#1\endcsname
\fi
#2}}
\providecommand{\BIBdecl}{\relax}
\BIBdecl

\bibitem{Riahi2012Finding}
F.~Riahi, Z.~Zolaktaf, M.~Shafiei, and E.~Milios, ``Finding expert users in
  community question answering,'' \emph{Topic Models Expert Recommender}, pp.
  791--798, 2012.

\bibitem{Zhao2016Expert}
Z.~Zhao, Q.~Yang, D.~Cai, X.~He, and Y.~Zhuang, ``Expert finding for
  community-based question answering via ranking metric network learning,'' in
  \emph{International Joint Conference on Artificial Intelligence}, 2016, pp.
  3000--3006.

\bibitem{Han2016Distributed}
F.~Han, S.~Tan, H.~Sun, M.~Srivatsa, D.~Cai, and X.~Yan, ``Distributed
  representations of expertise,'' in \emph{Siam International Conference on
  Data Mining}, 2016, pp. 531--539.

\bibitem{Balog2012Expertise}
K.~Balog, Y.~Fang, M.~De~Rijke, P.~Serdyukov, and L.~Si, ``Expertise
  retrieval,'' \emph{Foundations and Trends in Information Retrieval}, vol.~6,
  no.~23, pp. 127--256, 2012.

\bibitem{Liu2005Finding}
X.~Liu, M.~Koll, and M.~Koll, ``Finding experts in community-based
  question-answering services,'' in \emph{ACM International Conference on
  Information and Knowledge Management}, 2005, pp. 315--316.

\bibitem{Lin2017A}
S.~Lin, W.~Hong, D.~Wang, and T.~Li, ``A survey on expert finding techniques,''
  \emph{Journal of Intelligent Information Systems}, pp. 1--25, 2017.

\bibitem{Zhang2007Expert}
J.~Zhang, J.~Tang, and J.~Li, \emph{Expert Finding in a Social Network}.\hskip
  1em plus 0.5em minus 0.4em\relax Springer Berlin Heidelberg, 2007.

\bibitem{Rode2017Entity}
H.~Rode, P.~Serdyukov, D.~Hiemstra, and H.~Zaragoza, ``Entity ranking on
  graphs: Studies on expert finding,'' \emph{Centre for Telematics and
  Information Technology University of Twente}, 2017.

\bibitem{Rafiei2015A}
M.~Rafiei and A.~A. Kardan, ``A novel method for expert finding in online
  communities based on concept map and pagerank,'' \emph{Human-centric
  Computing and Information Sciences}, vol.~5, no.~1, pp. 1--18, 2015.

\bibitem{Boeva2017Data}
V.~Boeva, M.~Angelova, and E.~Tsiporkova, ``Data-driven techniques for expert
  finding,'' in \emph{International Conference on Agents and Artificial
  Intelligence}, 2017, pp. 535--542.

\bibitem{Wang2013ExpertRank}
G.~A. Wang, J.~Jiao, A.~S. Abrahams, W.~Fan, and Z.~Zhang, ``Expertrank: A
  topic-aware expert finding algorithm for online knowledge communities,''
  \emph{Decision Support Systems}, vol.~54, no.~3, pp. 1442--1451, 2013.

\bibitem{Dargahi2017Skill}
A.~Dargahi~Nobari, S.~Sotudeh~Gharebagh, and M.~Neshati, ``Skill translation
  models in expert finding,'' in \emph{International ACM SIGIR Conference on
  Research and Development in Information Retrieval}, 2017, pp. 1057--1060.

\bibitem{Li2015Social}
Y.~Li, S.~Ma, and R.~Huang, ``Social context analysis for topic-specific expert
  finding in online learning communities,'' 2015.

\bibitem{Zhou2012Topic}
G.~Zhou, S.~Lai, K.~Liu, and J.~Zhao, ``Topic-sensitive probabilistic model for
  expert finding in question answer communities,'' in \emph{ACM International
  Conference on Information and Knowledge Management}, 2012, pp. 1662--1666.

\bibitem{Liu2013Integrating}
D.~R. Liu, Y.~H. Chen, W.~C. Kao, and H.~W. Wang, ``Integrating expert profile,
  reputation and link analysis for expert finding in question-answering
  websites,'' \emph{Information Processing and Management An International
  Journal}, vol.~49, no.~1, pp. 312--329, 2013.

\bibitem{Yeniterzi2014Constructing}
R.~Yeniterzi and J.~Callan, ``Constructing effective and efficient
  topic-specific authority networks for expert finding in social media,'' pp.
  45--50, 2014.

\bibitem{Zhu2014Ranking}
H.~Zhu, E.~Chen, H.~Xiong, H.~Cao, and J.~Tian, ``Ranking user authority with
  relevant knowledge categories for expert finding,'' \emph{World Wide
  Web-internet and Web Information Systems}, vol.~17, no.~5, pp. 1081--1107,
  2014.

\bibitem{Bouguessa2008Identifying}
M.~Bouguessa and S.~Wang, ``Identifying authoritative actors in
  question-answering forums: the case of yahoo! answers,'' in \emph{ACM SIGKDD
  International Conference on Knowledge Discovery and Data Mining}, 2008, pp.
  866--874.

\bibitem{Liu2011Competition}
J.~Liu, Y.~I. Song, and C.~Y. Lin, ``Competition-based user expertise score
  estimation.'' in \emph{Proceeding of the International ACM SIGIR Conference
  on Research and Development in Information Retrieval, SIGIR 2011, Beijing,
  China, July}, 2011, pp. 425--434.

\bibitem{Zhou2014An}
G.~Zhou, J.~Zhao, T.~He, and W.~Wu, ``An empirical study of topic-sensitive
  probabilistic model for expert finding in question answer communities,''
  \emph{Knowledge-Based Systems}, vol.~66, no.~9, pp. 136--145, 2014.

\bibitem{Zhang2008A}
J.~Zhang, J.~Tang, L.~Liu, and J.~Li, ``A mixture model for expert finding,''
  \emph{Lecture Notes in Computer Science}, vol. 5012, pp. 466--478, 2008.

\bibitem{Deng2009Formal}
H.~Deng, I.~King, and M.~R. Lyu, ``Formal models for expert finding on dblp
  bibliography data.'' in \emph{Eighth IEEE International Conference on Data
  Mining}, 2009, pp. 163--172.

\bibitem{Daud2010Temporal}
A.~Daud, J.~Li, L.~Zhou, and F.~Muhammad, ``Temporal expert finding through
  generalized time topic modeling,'' \emph{Knowledge-Based Systems}, vol.~23,
  no.~6, pp. 615--625, 2010.

\bibitem{Topicmodeling2013Topic}
S.~Momtazi and F.~Naumann, ``Topic modeling for expert finding using latent
  dirichlet allocation,'' \emph{Wiley Interdisciplinary Reviews Data Mining and
  Knowledge Discovery}, vol.~3, no.~5, p. 346¨C353, 2013.

\bibitem{Liu2013An}
J.~Liu, L.~I. Qi, B.~Liu, and Y.~Zhang, ``An expert finding method based on
  topic model,'' \emph{Journal of National University of Defense Technology},
  vol.~35, no.~2, pp. 127--131, 2013.

\bibitem{Lin2013Finding}
L.~Lin, Z.~Xu, Y.~Ding, and X.~Liu, ``Finding topic-level experts in scholarly
  networks,'' \emph{Scientometrics}, vol.~97, no.~3, pp. 797--819, 2013.

\bibitem{Hashemi2013Expertise}
S.~H. Hashemi, M.~Neshati, and H.~Beigy, ``Expertise retrieval in bibliographic
  network: a topic dominance learning approach,'' pp. 1117--1126, 2013.

\bibitem{Yang2013CQArank}
Yang, Liu, Qiu, Minghui, Gottipati, Swapna, Zhu, Feida, Jiang, and Jing,
  ``Cqarank: jointly model topics and expertise in community question
  answering,'' 2013.

\bibitem{Wei2017Collaborative}
J.~Wei, J.~He, K.~Chen, Y.~Zhou, and Z.~Tang, ``Collaborative filtering and
  deep learning based recommendation system for cold start items,''
  \emph{Expert Systems with Applications}, vol.~69, pp. 29--39, 2017.

\bibitem{Li2017Deep}
Q.~Li and X.~Zheng, ``Deep collaborative autoencoder for recommender systems: A
  unified framework for explicit and implicit feedback,'' 2017.

\bibitem{Ying2016Collaborative}
H.~Ying, L.~Chen, Y.~Xiong, and J.~Wu, ``Collaborative deep ranking: A hybrid
  pair-wise recommendation algorithm with implicit feedback,'' in
  \emph{Pacific-Asia Conference on Knowledge Discovery and Data Mining}, 2016,
  pp. 555--567.

\bibitem{Rani2015Expert}
S.~K. Rani, K.~Raju, and V.~V. Kumari, ``Expert finding system using latent
  effort ranking in academic social networks,'' \emph{International Journal of
  Information Technology and Computer Science}, vol.~7, no.~2, pp. 21--27,
  2015.

\bibitem{Karimzadehgan2009Enhancing}
M.~Karimzadehgan, R.~W. White, and M.~Richardson, ``Enhancing expert finding
  using organizational hierarchies,'' in \emph{European Conference on Ir
  Research on Advances in Information Retrieval}, 2009, pp. 177--188.

\bibitem{DawitYimam2003Expert}
DawitYimam-Seid and AlfredKobsa, ``Expert-finding systems for organizations:
  Problem and domain analysis and the demoir approach,'' \emph{Journal of
  Organizational Computing}, vol.~13, no.~1, pp. 1--24, 2003.

\bibitem{Bozzon2013Choosing}
A.~Bozzon, M.~Brambilla, S.~Ceri, M.~Silvestri, and G.~Vesci, ``Choosing the
  right crowd: expert finding in social networks,'' in \emph{International
  Conference on Extending Database Technology}, 2013, pp. 637--648.

\bibitem{Kardan2011Expert}
A.~Kardan, A.~Omidvar, and F.~Farahmandnia, ``Expert finding on social network
  with link analysis approach,'' in \emph{Electrical Engineering}, 2011, pp.
  1--6.

\bibitem{Li2013A}
X.~Li, J.~Ma, Y.~Yang, and D.~Wang, ``A service mode of expert finding in
  social network,'' in \emph{International Conference on Service Sciences},
  2013, pp. 220--223.

\bibitem{Cheng2015Exploiting}
X.~Cheng, S.~Zhu, G.~Chen, and S.~Su, ``Exploiting user feedback for expert
  finding in community question answering,'' in \emph{IEEE International
  Conference on Data Mining Workshop}, 2015, pp. 295--302.

\bibitem{Mimno2007Expertise}
D.~Mimno and A.~Mccallum, ``Expertise modeling for matching papers with
  reviewers,'' in \emph{ACM SIGKDD International Conference on Knowledge
  Discovery and Data Mining, San Jose, California, Usa, August}, 2007, pp.
  500--509.

\bibitem{Liang2016Formal}
S.~Liang and M.~D. Rijke, \emph{Formal language models for finding groups of
  experts}.\hskip 1em plus 0.5em minus 0.4em\relax Pergamon Press, Inc., 2016.

\bibitem{Alarfaj2013Finding}
F.~Alarfaj, U.~Kruschwitz, D.~Hunter, and C.~Fox, ``Finding the right
  supervisor: expert-finding in a university domain,'' pp. 1--6, 2013.

\bibitem{Li2015A}
H.~Li, S.~Jin, and L.~Shudong, ``A hybrid model for experts finding in
  community question answering,'' in \emph{International Conference on
  Cyber-Enabled Distributed Computing and Knowledge Discovery}, 2015, pp.
  176--185.

\bibitem{On2017Dynamicity}
H.~B. Mahmood~Neshati, Zohreh~Fallahnejad, ``On dynamicity of expert finding in
  community question answering,'' pp. 1026--1042, 2017.

\bibitem{Liu2015ZhihuRank}
X.~Liu, S.~Ye, X.~Li, Y.~Luo, and Y.~Rao, ``Zhihurank: A topic-sensitive expert
  finding algorithm in community question answering websites,'' 2015.

\bibitem{Zhao2015Cold}
Z.~Zhao, F.~Wei, M.~Zhou, and W.~Ng, \emph{Cold-Start Expert Finding in
  Community Question Answering via Graph Regularization}.\hskip 1em plus 0.5em
  minus 0.4em\relax Springer International Publishing, 2015.

\bibitem{koren2009mf}
Y.~Koren, R.~Bell, C.~Volinsky \emph{et~al.}, ``Matrix factorization techniques
  for recommender systems,'' \emph{Computer}, vol.~42, no.~8, pp. 30--37, 2009.

\bibitem{Linden2003Amazon}
G.~Linden, B.~Smith, and J.~York, ``Amazon.com recommendations: Item-to-item
  collaborative filtering,'' \emph{IEEE Internet Computing}, vol.~7, no.~1, pp.
  76--80, 2003.

\bibitem{Koren2008FM}
Y.~Koren, ``Factorization meets the neighborhood: a multifaceted collaborative
  filtering model,'' in \emph{Proceedings of the 14th ACM SIGKDD international
  conference on Knowledge discovery and data mining}.\hskip 1em plus 0.5em
  minus 0.4em\relax ACM, 2008, pp. 426--434.

\bibitem{Dai2016Recurrent}
H.~Dai, Y.~Wang, R.~Trivedi, and L.~Song, ``Recurrent coevolutionary feature
  embedding processes for recommendation,'' 2016.

\bibitem{Rendle2011FM}
S.~Rendle, ``Factorization machines,'' \emph{IEEE International Conference on
  Data Mining}, pp. 995--1000, 2011.

\bibitem{Rendle2012libFM}
------, ``Factorization machines with {libFM},'' \emph{ACM Trans. Intell. Syst.
  Technol.}, vol.~3, no.~3, pp. 57:1--57:22, 2012.

\bibitem{friedman2001}
J.~H. Friedman, ``Greedy function approximation: A gradient boosting machine,''
  \emph{Annals of Statistics}, vol.~29, no.~5, pp. 1189--1232, 2001.

\bibitem{XgboostPPT}
T.~Chen, ``{I}ntroduction to {B}oosted {T}rees,'' 2014,
  \url{http://homes.cs.washington.edu/~tqchen/pdf/BoostedTree.pdf}.

\bibitem{Xgboost}
T.~Chen and C.~Guestrin, ``Xgboost: A scalable tree boosting system,''
  \emph{arXiv preprint arXiv:1603.02754}, 2016.

\bibitem{Autorec2015}
S.~Sedhain, A.~K. Menon, S.~Sanner, and L.~Xie, ``Autorec: Autoencoders meet
  collaborative filtering,'' in \emph{Proceedings of the 24th International
  Conference on World Wide Web}.\hskip 1em plus 0.5em minus 0.4em\relax ACM,
  2015, pp. 111--112.

\bibitem{CFNADE}
Y.~Zheng, B.~Tang, W.~Ding, and H.~Zhou, ``A neural autoregressive approach to
  collaborative filtering,'' \emph{arXiv preprint arXiv:1605.09477}, 2016.

\bibitem{Match-SRNN}
S.~Wan, Y.~Lan, J.~Guo, J.~Xu, L.~Pang, and X.~Cheng, ``Match-srnn: Modeling
  the recursive matching structure with spatial rnn,'' in \emph{Proceedings of
  the 25th International Joint Conference on Artificial Intelligence (IJCAI)},
  2016.

\bibitem{joac2006}
T.~Joachims, ``Training linear svms in linear time,'' in \emph{Proceedings of
  the 12th ACM SIGKDD international conference on Knowledge discovery and data
  mining}.\hskip 1em plus 0.5em minus 0.4em\relax ACM, 2006, pp. 217--226.

\bibitem{zhou2012}
Z.~H. Zhou, \emph{Ensemble Methods: Foundations and Algorithms}.\hskip 1em plus
  0.5em minus 0.4em\relax CRC Press, 2012.

\bibitem{Breiman1996}
L.~Breiman, ``Bagging predictors,'' \emph{Machine Learning}, vol.~26, no.~2,
  pp. 123--140, 1996.

\bibitem{Friedman2000Additive}
J.~Friedman, T.~Hastie, and R.~Tibshirani, ``Additive logistic regression: a
  statistical view of boosting,'' \emph{Annals of Statistics}, vol.~28, no.~2,
  pp. 337--374, 2000.

\bibitem{Johnson2001Bootstrap}
R.~W. Johnson, ``An introduction to the bootstrap,'' \emph{Teaching
  Statistics}, vol.~23, no.~2, p. 49¨C54, 2001.

\bibitem{SVDFeature}
\BIBentryALTinterwordspacing
T.~Chen, W.~Zhang, Q.~Lu, K.~Chen, Z.~Zheng, and Y.~Yu, ``{SVDFeature}: A
  toolkit for feature-based collaborative filtering,'' \emph{Journal of Machine
  Learning Research}, vol.~13, pp. 3619--3622, 2012. [Online]. Available:
  \url{http://www.jmlr.org/papers/volume13/chen12a/chen12a.pdf}
\BIBentrySTDinterwordspacing

\bibitem{tsne2017}
N.~Pezzotti, B.~Lelieveldt, L.~V.~D. Maaten, T.~Hollt, E.~Eisemann, and
  A.~Vilanova, ``Approximated and user steerable tsne for progressive visual
  analytics,'' \emph{IEEE Trans Vis Comput Graph}, vol.~PP, no.~99, pp. 1--1,
  2017.

\bibitem{nmf1994}
P.~Paatero and U.~Tapper, ``Positive matrix factorization: A nonnegative factor
  model with optimal utilization of error estimates of data values,''
  \emph{Environmetrics}, vol.~5, no.~2, pp. 111--126, 1994.

\bibitem{Adam2014}
D.~P. Kingma and J.~Ba, ``Adam: A method for stochastic optimization,''
  \emph{Computer Science}, 2014.

\bibitem{deepwalk2014}
B.~Perozzi, R.~Al-Rfou, and S.~Skiena, ``Deepwalk: online learning of social
  representations,'' pp. 701--710, 2014.

\end{thebibliography}

\end{document}